\documentclass{article}

\usepackage[utf8]{inputenc} % allow utf-8 input
\usepackage[T1]{fontenc}    % use 8-bit T1 fonts
\usepackage{graphicx}
\usepackage{booktabs}
\usepackage[table]{xcolor}
\usepackage{multirow}
\usepackage{amsmath,amsfonts,amssymb}
\usepackage{optidef}
\usepackage{bbm,bm}
\usepackage{subfig}
\usepackage{parskip}
\usepackage[linesnumbered,ruled,vlined]{algorithm2e}
\usepackage[numbers]{natbib}
\usepackage{afterpage}

\graphicspath{ {./images/} {./images/plots/} {./images/plots/carr/}}
\newcommand \W{\mathbf{w}}

\DeclareMathOperator*{\argmin}{arg\,min}
\newcommand\E{\mathbb{E}}

\newcommand\R{\mathbb{R}}

\title{Data-Driven Approach for Static Hedging of Exchange-Traded Index Options: Empirical Insights and Comparative Analysis with Dynamic Hedging in the NSE Market}
\author{
Vikranth Lokeshwar Dhandapani\footnote{Corresponding author. Email: \texttt{vikranthl@iisc.ac.in}}, Shashi Jain \\
  Department of Management Studies, \\
  Indian Institute of Science, \\
  Bangalore.\\
}

\begin{document}
%%%%%%%%%%%%%%%%
\maketitle
\begin{abstract}
This paper presents a data-driven interpretable machine learning algorithm for semi-static hedging of Exchange Traded options, considering transaction costs with efficient run-time. Further, we provide empirical evidence on the performance of hedging longer-term National Stock Exchange (NSE) Index options using a self-replicating portfolio of shorter-term options and cash position, achieved by the automated algorithm, under different modeling assumptions and market conditions, including Covid period. We also systematically assess the model's performance using the Superior Predictive Ability (SPA) test by benchmarking against the static hedge proposed by \citet{carr2014static} and industry-standard dynamic hedging. We finally perform a thorough Profit and Loss (PnL) attribution analysis on the target option and hedge portfolios (dynamic and static) to discern the factors explaining the superior performance of static hedging. 
\end{abstract}

%% keywords can be removed
%\keywords{Static Hedging \and Lasso Regression \and Dynamic Hedging \and Superior Predictive Ability \and Profit and Loss Attribution Analysis}

\section{Introduction}

Hedging is a widely practiced risk management tool to reduce the risk due to adverse price movements of options. Dynamic and static hedging are two common approaches to hedge options. Dynamic hedging involves continuously adjusting the position to maintain a delta-neutral position. In contrast, static hedging involves using a pre-determined portfolio to replicate the option's pay-off regardless of the market conditions or changes in the underlying instrument hedged. 

Numerous studies have explored diverse approaches to both dynamic and static hedging. However, there is a noticeable scarcity of research that rigorously compares their relative performance empirically. This paper makes the following contributions, firstly, it introduces a data-driven framework for semi-static hedging of Exchange-traded options, taking into account transaction costs. Additionally, the study centers on empirically evaluating the performance of hedging longer-term index options traded on the National Stock Exchange (NSE) in India, employing shorter-term options on the same index. Our approach incorporates real-time trading constraints, including the availability and liquidity of shorter-term options used in constructing the hedge, as well as transaction costs.  We compare the hedging efficiency of our constructed static hedge with both dynamic hedging and the Carr-Wu static hedge (\citet{carr2014static}). Identifying the best model becomes a non-standard inference problem, involving the comparison of historical performances across various hedging approaches. To address the possibility that the best model might have lower sample performance than some inferior models when comparing multiple models, we employ the Superior Predictive Ability (SPA) test developed by \citet{hansen2005test}. This approach facilitates multiple comparisons to determine whether other models significantly outperform the benchmark model and helps identify the best model with minimal statistical hedging error. We conduct these comparisons across different market periods, underlying indices, option types, and moneyness levels. Our empirical findings support the superiority of the proposed static hedging model over the dynamic hedging model, especially during periods of heightened market volatility or substantial underlying asset jumps. Additionally, we conduct a detailed Profit and Loss (PnL) attribution analysis to discern the factors contributing to the exceptional performance of the proposed static hedging model. Our analysis demonstrates that this superiority arises from the static hedge's ability to address both delta and gamma risks, in contrast to delta hedging, which only mitigates delta risk.

 Dynamic delta hedging requires a daily rebalancing of the hedge portfolio by taking positions on the underlying or the futures. In an ideal setup, hedging is instantaneous; however, due to impractical transaction costs, discrete rebalancing is adopted, which may give rise to hedge error. Further, it is a known deficiency of dynamic hedging that large movements in the underlying and highly volatile conditions may cause significant losses. We have been experiencing extreme market events recently, such as the 2007 world crisis and Covid market conditions. \citet{bakshi2003delta} demonstrated that the delta hedge strategy underperforms at times of higher volatility. The effectiveness of the hedge is highly dependent on the frequency of rebalancing, and the transaction costs can significantly reduce the performance of the hedge. Many works address the drawbacks of dynamic hedging - essentially to create more robust dynamic hedging, especially for complex pay-offs with barriers or early exercise features and extreme market conditions of high volatility, jumps, and illiquidity. Some notable literature on robust hedging includes \citet{brown2001robust} for barrier options, \citet{hobson1998robust} for lookback options, \citet{fabozzi2016improved} for American options, \citet{wilmott2000feedback} for illiquid markets,  \citet{he2006calibration} under jump diffusion and recently, \citet{lutkebohmert2022robust} for robust deep hedging under parameter incertainty. \citet{coleman2003dynamic} showed that the performance of dynamic hedging with local volatility function (computed as proposed by \citet{coleman2001reconstructing}) had a smaller hedge error compared to dynamic hedging with implied volatility. \citet{kennedy2009dynamic} set up a dynamic hedging strategy using a hedge portfolio consisting of the underlying asset and liquidly traded options to minimize the hedging errors in the presence of transaction cost under jump-diffusion frameworks. In addition, there are considerable efforts to devise alternative hedging strategies, such as static hedging. 

 Static hedging is a risk management technique used to reduce the exposure of options to market fluctuations by constructing a pre-determined portfolio that replicates the pay-offs of the option (especially with complex pay-offs) to be hedged. \citet{breeden1978prices} pioneered an alternative approach to match pay-offs instead of risk sensitivities, which is robust by design as contracts with matching pay-offs will behave similarly regardless of underlying risk dynamics even with the presence of random jumps. The fundamental idea is that a portfolio of standard options with the same maturity as the claim can replicate a path-independent claim. However, the class of claims that this approach can hedge is limited and cannot hedge options with different maturities. The idea was further elaborated by \citet{green1987spanning}, \citet{nachman1988spanning} and \citet{carr2001optimal}. \citet{carr1998static} discusses the problem of hedging exotic, including barrier and lookback options with static portfolios of standard options. \citet{takahashi2009new} proposed a new scheme for static hedging of European path-independent derivatives under stochastic volatility models by applying the static replication method for one-dimensional price processes developed by \citet{takahashi2007efficient}. \citet{carr2009put} relaxes the assumptions and extends the put-call symmetry in several directions, including unified local/stochastic volatility models and asymmetric dynamics. Subsequently, a conjugate European-style claim of equal value is constructed for any given European-style or barrier-style pay-off, enabling a semi-static hedge. \citet{carr2011static} considered the problem of hedging a barrier option with a semi-static position in a European-type contingent claim of the same maturity. 

 \citet{carr2014static} (referred to as Carr-Wu static hedge in this paper) derived a new static spanning relation between a target option and a continuum of shorter-term options written on the same asset under a single-factor Markovian setting. Further, they developed an approximation for the static hedging strategy using only a finite number of short-term options with the strike levels and the associated portfolio weights obtained using a Gauss–Hermite quadrature method. However, in the real trading scenario, the strikes obtained by the Gauss-Hermite method need not necessarily be available or liquid. \citet{wu2016simple} provides a new hedging strategy of the target option using three options at different maturities and strikes based on approximate matching of target option function expansion along with maturity and strike rather than risk factors. \citet{leung2016optimal} presents a flexible framework for hedging a contingent claim through static positions in vanilla European options, bonds, and forwards. The optimal static hedging strategy is derived using a model-free expression, which minimizes the expected squared hedging error and is subject to a cost constraint. To illustrate the versatility of this approach, the author presents several numerical examples in the paper. \citet{bossu2021functional} showed that the spectral decomposition techniques might achieve exact pay-off replication of European Options with a discrete portfolio of special options. Further, \citet{bossu2022static} extended the study on static replication to European standard dispersion options written on the Euclidean norm of a vector of multi-asset performances with the help of integral equation techniques. 

 The study on the application of machine learning in hedging has also increased over the last decade, particularly explainable frameworks. \citet{buehler2019deep} presents a framework for using deep reinforcement learning to hedge derivative portfolios in the presence of market frictions. Additionally, the authors show that their algorithm can approximate any optimal solution using constrained trading strategies, making it broadly applicable across different market dynamics and hedging instruments. Finally, they demonstrate the effectiveness of their approach using a synthetic market driven by the Heston model, outperforming the standard complete market solution in hedging under transaction costs. \citet{lokeshwar2022explainable} presented an explainable neural network for pricing and static hedging of OTC options by learning the weights and strikes of the hedging portfolio.

\section{Static Hedging Framework}
\label{Static Hedging Framework}

This section illustrates the static hedging framework and introduces the notations used in this paper. We are interested in hedging a long-term European option maturing at $T_2$ by a self-replicating portfolio of liquid exchange traded European options with a shorter maturity $T_1$ ($T_2 \ge T_1$).  We call the long term option to be hedged as the target option and the portfolio of short maturity option as the hedge portfolio.  The hedge also contains a cash position $C(t)$ at any time $t$, such that, $T_0=0 \le t  \le T_1$. Let the weights of the constituent options be $\mathbf{w} = [w_1, w_2, .... w_p]^\intercal \in \mathbb{R}^{p}$, where, $p$ is the number of liquid options used in the hedge portfolio. We propose a Monte-Carlo-based machine learning algorithm to learn $\W$ and the cash to be held in the hedge portfolio.

We assume a complete probability space $(\Omega,\mathcal{F},\mathbb{P})$ and finite time horizon $[0,T_2]$. $\Omega$ is the set of all possible realizations of the stochastic economy between 0 and $T_2$. 
The information structure in this economy is represented by an augmented filtration $\mathcal{F}_t:t\in[0,T_2],$  with
$\mathcal{F}_t$ the sigma field of distinguishable events at time $t$, and $\mathbb{P}$ is the risk-neutral probability measure on elements of $\mathcal{F}.$  It is assumed that $\mathcal{F}_t$ is generated by $W_t$, a standard Brownian motion, and the state of economy is represented by an $\mathcal{F}_t$-adapted Markovian process, $S_t \in \mathbb{R}$, which has dependence on model parameters $\Theta = \{\theta_1,\ldots,\theta_{N_{\theta}}\}.$

We assume that the underlying asset satisfies the following stochastic differential equation:

\begin{equation}\label{masterprocess}
\frac{dS_t}{S_t} = (r')dt + \sigma_t dW_t, 
 \end{equation}

where  $r^{'}$ is the stock or the index return implied from the futures market (in other words, $r^{'} = r - q$, $r$ is the risk-free interest rate and $q$ is the implied dividend yield for the underlying asset). The instantaneous volatility at time $t,$  $\sigma_t,$ maybe random, and $W_{t}$ is a Brownian Motion. 

We use $V_t(K,T)$ to denote the time $t$ price of a European option with strike $K$ and maturity $T.$ We assume that there exists a pricing function,
\begin{equation}\label{VDefine}
V_t(K, T) := V(S_t, t; K, T; \Theta), t \in [0, T], K \ge 0.
\end{equation}

Let $K^{*} \in \mathbb{R}$ be the strike of the target option maturing at $T_2,$ and 
$$V(S_{T_1}(\omega), T_1;K^*, T_2),$$
be the target option price at future time $T_1$ for a stochastic realisation $S_{T_1}(\omega),$ where $\omega \in \Omega.$  The strikes of liquid options in the hedging portfolio maturing at $T_1$ be $\mathbf{K} = (\kappa_{i})_{i=1,\ldots, p}^{\intercal} \in \mathbb{R}^{p}$. We denote the option payoff for each shorter-term option considered as 
$$\phi_{\kappa_i}(S_{T_1}(\omega)) = \max\left( i_{cp} \cdot (S_{T_1}(\omega) - \kappa_i), 0 \right),$$

 where, $i_{cp} = +1$ for call options and $-1$ for put options. The payoff vector of the options in the hedge portfolio corresponding to a stochastic realisation $S_{T_1}(\omega), \, \omega \in \Omega$ at $T_1$ is represented as 
 \begin{equation}\label{phiDefine}
 \bm{\phi}(S_{T_1}) := \left[ \phi_{\kappa_1} \left(S_{T_1}(\omega)\right),\ldots,\phi_{\kappa_p}\left(S_{T_1}(\omega)\right) \right] \in \mathbb{R}^{1 \times (p)}.
 \end{equation}

We are interested in learning the weights, $\W,$ of the shorter maturity options in the hedge portfolio and the cash position $C(T_1)$ such that the hedge portfolio replicates the target option price at time $T_1.$ Under no arbitrage and Markovian assumption this would imply the hedge portfolio would replicate the target option at any time $t \le T_1.$

\subsection{The Carr-Wu approach for static hedging }
\label{Carr_Wu Section}

  In their work, \citet{carr2014static}  demonstrate how a European option can be replicated using a continuum of shorter-maturity European options in a continuous time one-factor Markovian setting. The portfolio's weights remain constant and do not vary with changes in the underlying security price or time, making it a static hedge. More precisely they show that the time $t$ value of a European call option maturing at $T_2 \ge t$ relates to the time $t$ value of a continuum of European call options at a shorter maturity $T_1 \in [t, T_2]$ by

\begin{equation}\label{carrWuEqn}
V(S,t;K^*,T_2;\Theta)=\int_0^{\infty} w(\mathcal{K})V(S,t;\mathcal{K},T_1;\Theta) d\mathcal{K}, S>0.
\end{equation}

The weights $w(\mathcal{K})$ do not vary with $S,$ or $t,$ and are given by

$$
w(\mathcal{K}) = \frac{\partial^2}{\partial \mathcal{K}^2} V(\mathcal{K},T_1;K^*,T_2;\Theta).
$$

As it is not practical to hold a continuum of securities, \citet{carr2014static} propose to approximate the spanning integral in Equation \ref{carrWuEqn} by a weighted sum of a finite number, $p,$  call options with strikes at $\mathcal{K}_j, j = 1, 2, \ldots,p,$

$$
\int_0^{\infty} w(\mathcal{K})V(S,t;\mathcal{K},T_1;\Theta) d\mathcal{K} \approx \sum_{j=1}^p \mathcal{W}_jV(S,t;\mathcal{K}_j, T_1;\Theta),
$$

where the strike points $\mathcal{K}_j$ and their corresponding weights are chosen based on the Gauss-Hermite quadrature rule. If liquid exchange traded securities are exclusively used to establish the hedge, only a limited number of strike prices based on the Gauss-Hermite quadrature rule will be accessible, which could result in an incomplete replication.

\subsection{The Neural Network based static hedging}\label{nnSummary}

Universal approximation theorem shows that any continuous function defined over a compact set can be approximated arbitrarily well using a sufficiently large number of neurons. \citet{lokeshwar2022explainable} use a neural network with a single hidden layer and ReLU activation function to replicate the target option value (written on potentially a $d$-dimensional underlying)  at forward time $T_1.$ 
This feed-forward network $\tilde{G}^{\beta} : \R^d \rightarrow\R$ is chosen as, 

\begin{equation*}
\tilde{G}^{\beta_{T_1}} := \psi \circ A_2 \circ \varphi \circ A_1
\end{equation*}
where $ A_1:\R^d \rightarrow \R^p$ and $ A_2:\R^p\rightarrow\R$ are affine functions of the form,

\begin{equation}
A_1(\mathbf{x}) = \mathbf{W}_1\mathbf{x} + \mathbf{b}_1 \ \ \textrm{for} \ \mathbf{x} \in \R^d,\ \mathbf{W}_1 \in \R^{p \times d}, \mathbf{b}_1 \in \R^p,
\end{equation}
and
\begin{equation}\label{weightEqn}
A_2(\mathbf{x}) = \mathbf{W}_2\mathbf{x} + b_2 \ \ \textrm{for} \ \mathbf{x} \in \R^p,\ \mathbf{W}_2 \in \R^{1 \times p}, b_2 \in \R.
\end{equation}

$\varphi : \R^j \rightarrow \R^j, j \in \mathbb{N}$ is the component-wise ReLU activation function given by:
\begin{equation*}
\varphi(x_1,\ldots,x_j):=\left(\max(x_1,0),\ldots,\max(x_j,0)\right),
\end{equation*}

while $\psi:\R^j \rightarrow \R^j, j \in \mathbb{N}$ is the component-wise linear activation function given by:
\begin{equation*}
\psi(x_1,\ldots,x_j):=\left(x_1,\ldots,x_j\right).
\end{equation*}

Specifically, $\beta_{T_{1}}\in \R^{N_p},$ where $N_p$ is the dimension of the parameter space ($N_p = 1+p+p+p \times d$)
is chosen to minimize the following mean squared error,

\begin{equation}\label{neuralOpt}
\beta_{T_1} = \argmin_{\beta_{T_1}}\E \left[ \left( V\left(S_{T_1}, T_1;K^*, T_2 ; \Theta\right)-\tilde{G}^{\beta_{T_1}} \left(S_{T_1}\right)\right)^2 \right].
\end{equation}

When the risk-neutral evolution of the stock price is Markov in the stock price $S$ and the calendar time $t,$ and when sufficiently large number of neurons, $p,$  are used, universal approximation theorem guarantees;

$$
V\left(S_{T_1}, T_1;K^*, T_2; \Theta \right) \approx \tilde{G}^{\beta_{T_1}} \left(S_{T_1}\right),
$$

or more precisely,

\begin{equation}\label{uat}
\left|V\left(S_{T_1}, T_1;K^*, T_2; \Theta\right) - \tilde{G}^{\beta_{T_1}} \left(S_{T_1}\right)\right| \le \epsilon,
\end{equation}

where $\epsilon$ can be chosen to be arbitrarily small. According to \citet{lokeshwar2022explainable} , $\tilde{G}^{\beta_{T_1}}$ can be understood as a portfolio of $p$ European options written on $S$ expiring at $T_1$. The parameters of the network $\tilde{G}^{\beta_{T_1}}$ are estimated using the stochastic gradient descent, which then determines the strikes and the weights of the option portfolio. 
The method proposed in the article by \citet{lokeshwar2022explainable} offers a feasible numerical approach to semi-statically replicate high-dimensional options. However, the shorter-term options they generate may not exactly match the available options on the market.

\subsection{The LASSO based static hedge}

We consider a problem where the option trader is interested in creating a semi-static hedge for the target option using only liquid exchange traded call and put options. The strikes of liquid call and put options expiring at $T_1$ are given by $\mathbf{K} = (\kappa_{i})_{i=1,\ldots, p}^{\intercal} \in \mathbb{R}^{p}$. Let $\mathcal{G}(S_t, t; \W, T_1)$ be the time $t$ value of a portfolio of options that mature at $T_1,$ when the corresponding  value of the underlying asset is $S_t.$ More precisely, 

$$
\mathcal{G}(S_t, t; \W, T_1) = \sum_{i=1}^p w_i V(S_t, t; \kappa_i, T_1; \Theta)
$$

The objective is to determine the weights $\W$ that minimize the expected squared replication error between the hedge portfolio $\mathcal{G}$ and the target option at $T_1,$ the expiry date for the short term options, i.e.

\begin{equation}\label{optProbEq}
\W = \argmin_{\W} \E\left[\left(V\left(S_{T_1}, T_1;K^*, T_2\right) - \mathcal{G}(S_{T_1}, T_1; \W, T_1)\right)^2\right],\, 0  \le T_1 <T_2.
\end{equation}

We base our analysis on the following essential assumptions: the underlying asset adheres to a Markovian process, the market operates without friction, and arbitrage opportunities are absent.

 In the case where the option is written on a single underlying asset, and only a fixed set of strikes is available in the market, the nodes in the hidden layer of the neural network proposed by \citet{lokeshwar2022explainable} are predetermined and consequently remain fixed. The sole parameter of interest for the network is the determination of the weight $\mathbf{W}_2,$ as specified in Equation \ref{weightEqn}. Due to the linearity of the neural network, $\tilde{G}^{\beta_{T_1}},$  with respect to $\mathbf{W}_2,$ the optimization problem presented in Equation \ref{neuralOpt} is equivalent to that in \ref{optProbEq}. A solution to Equation \ref{optProbEq} can be efficiently obtained through linear regression.  
 
 In this study, we propose the use of lasso regression (\citet{tibshirani1996regression}) as a numerical method for determining the composition of the hedging portfolio derived from solving the optimization problem in Equation \ref{optProbEq}. We considered various alternate regression models, including multiple linear regression, artificial neural networks, and ridge regression. However, the lasso regression stands out due to its regularization (see Equation \ref{eq:lasso_opt2}), which shrinks certain model coefficients and forces other insignificant coefficients to exactly zero. The financial implication of this property are noteworthy, as it allows us to disregard shorter-term options for which portfolio weights become zero during the hedging process. Consequently, we can achieve a hedging portfolio with fewer constituents. 
 
In situations where an arbitrarily large number of strikes is available at  \(T_1\), the solution to the LASSO-based static hedge is equivalent to the neural network based model proposed by \citet{lokeshwar2022explainable}). This allows the assertion that the absolute error between the target option and the LASSO-based static hedge can be arbitrarily well bounded, as stated in Equation \ref{uat}.

To learn the portfolio weights of the hedging portfolio and the cash position at time-zero, lasso regression is performed at simulation time $T_1$, where, the regressor variables vector is $\bm{\phi}(S_{T_1})$  and response variable is $V\left(S_{T_1}, T_1;K^*, T_2\right),$ as defined in Equation \ref{phiDefine}, and \ref{VDefine} respectively. 

Based on $N$ Monte-Carlo simulations of future index levels at $T_1$, i.e., $S_{T_1}(\omega_i)$, $i = 1,\ldots, N,$  the N replications of the generated data can be summarised as,  
$$ 
\left(X_{i} = \bm{\phi}\left(S_{T_1}(\omega_i) \right), \ y_{i} = V\left(S_{T_1}(\omega_i), T_1;K^*, T_2\right) \right), \,\, , i = 1, \ldots, N. 
$$ 
Further, we define,
$$
\textbf{X} = \begin{bmatrix}
1 & X_1 \\
\vdots& \vdots\\
1 & X_N
\end{bmatrix} \in \mathbb{R}^{N \times (p+1)},  \text{ and vector } \textbf{Y} =\begin{bmatrix}
y_1 \\
\vdots\\
y_N
\end{bmatrix} \in \mathbb{R}^{N}
$$

Let $\tilde{W} \in \mathbb{R}^{(p + 1) \times 1}$ be the corresponding vector of weights,  then the lasso optimisation problem can be written as,

\begin{mini}[2]                   % mini! = minimize 
    {\tilde{W} \in \mathbb{R}^{p+1}}                               % optimization variable
    {\frac{1}{N}  \ \Vert  {\textbf{X}} \tilde{W} \ - \ \textbf{Y} \Vert^{2}_{2} }   % objective function and label
    {\label{eq:lasso_opt}}             % label for optimizatio problem
    {}                                % optimization result
    \addConstraint{\Vert \tilde{W} \Vert_{1}}{\leq \zeta},   % constraint 1
%    \addConstraint{x}{\in X \label{eq:con2}}  % constraint 2
\end{mini}

 with the equivalent in Lagrangian form being, 

	\begin{mini}[2]                   % mini! = minimize 
	    {\tilde{W} \in \mathbb{R}^{p+1}}                               % optimization variable
	    {\frac{1}{N} \ \Vert  {\textbf{X}} \tilde{W} \ - \ \textbf{Y} \Vert^{2}_{2} + \lambda \Vert \tilde{W} \Vert_{1}} % objective function and label
	    {\label{eq:lasso_opt2}}             % label for optimizatio problem
	    {}  ,                             % optimization result
	\end{mini}

where $\Vert . \Vert_{1}$ and $\Vert . \Vert_{2}$ denote the L1 and L2 norms, respectively, $\zeta(\ge 0)$ serves as a tuning parameter, and $\lambda \ge 0$ is the Lagrange multiplier. The user-defined input $\lambda$ plays a crucial role in the optimization problem by acting as a shrinkage parameter, mitigating multi-collinearity, and reducing model complexity. When $\lambda = 0$, the lasso simplifies to multiple linear regression. As $\lambda$ approaches $\infty$, all coefficients converge to $0$. In essence, $\lambda$ facilitates the control of the trade-off between model complexity and accuracy.

The optimal $\lambda$ is determined by examining the point at which the learning curve (depicting $\lambda$ on the x-axis and model error on the y-axis) plateaus. Concerning the static hedge portfolio, $\tilde{W}$ represents the vector of weights allocated to the $p$ options maturing at $T_1,$ with the first weight corresponding to the amount invested in the risk-free asset.

A generalized  steps for performing a Lasso Static Hedge is provided as Algorithm \ref{algo}. In Section \ref{Empirical Analysis of Lasso Static Hedge} we discuss the details on the methodology, market data, and analysis of the lasso static hedge regression. Further, the model fit shown is based on the index simulations independent of that used to train the model. The absolute mean error as a ratio of underlying index levels is in the order of $10^{-5}$ in all the cases considered.   

\clearpage

\begin{algorithm}[!ht]
\DontPrintSemicolon

Generate Implied Volatility Surface as of time $T_0$ based on liquid options available with maturities corresponding to target longer-term options and shorter-term options of hedge portfolio.\\ 
 
Create the strikes vector $K$ by filtering all liquid shorter-term Call and Put options for creating the hedging portfolio \\

Generate $S_{T_1} (\omega_i)$ for $i=1, \ldots , 5000$ using GBM with index returns implied from futures market and at-the-money (ATM) volatility corresponding to the tenor $T_1 - T_0.$\\ 

Calculate target option price $V\left(S_{T_1}(\omega_i), T_1;K^*, T_2\right)$ based on the choice of model (with volatility obtained from Step 1), and obtain the hedging portfolio payoff vector $\bm{\phi}(S_{T_1}(\omega_i))$ for each simulation.  \\

Perform a Lasso Regression fit between $V\left(S_{T_1}(\omega_i), T_1;K^*, T_2\right)$ and $\bm{\phi}(S_{T_1}(\omega_i))$ to generate the shorter term hedge portfolio weights and the cash position.  \\

Repeat the hedging at the frequency of shorter-term maturity.

\caption{Static hedging strategy using lasso}
\label{algo}
\end{algorithm}

\afterpage{\clearpage}

\section{Empirical Analysis of Lasso Static Hedge}
\label{Empirical Analysis of Lasso Static Hedge}

\begin{figure}[hbt!]
\begin{center}
\includegraphics[width=4.5in, height=3.5in]{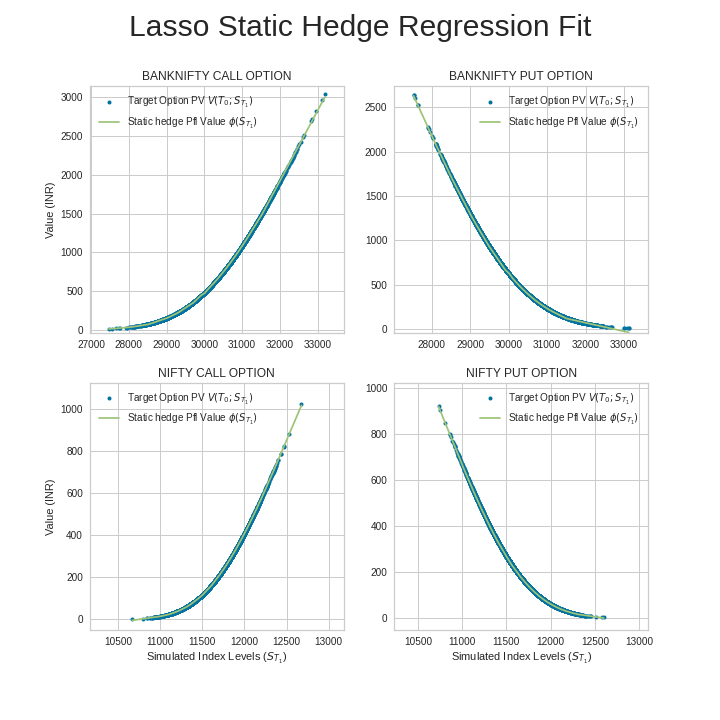}
\caption{\footnotesize Lasso regression model fit comparing the monthly target option value and weekly static hedge portfolio payoff (all constituent options expiring at $T_1$) across simulated index levels for call and put options on NIFTY and BANKNIFTY indices in the four subplots for an example snapshot date. The x-axis depicts simulated index levels at time $T_1$ based on information available at time $T_0$. The y-axis represents present values, where blue dots correspond to the target option value, and the green line represents the static hedge portfolio payoff expiring at $T_1$.} \label{lasso_regression_fit.png}
\end{center}
\end{figure}

When there is a large number of liquid strikes available for short maturity options, due to the equivalence of the lasso based regression with the neural network proposed in \citep{lokeshwar2022explainable}, one can fit $V\left(S_{T_1}(\omega_i), T_1;K^*, T_2\right)$ arbitrarily well using the payoff of the short term option portfolio maturity at $T_1.$ Figure \ref{lasso_regression_fit.png}  illustrates one such example, where simulated target option $V\left(S_{T_1}(\omega_i), T_1;K^*, T_2\right),$ at $T_1$ is replicated by the corresponding aggregated payoff of options maturing at $T_1.$ The weights of the short term options in the hedge portfolio are determined using the lasso regression.

This section empirically studies the performance of the proposed lasso regression-based static hedge performance for the options traded in National Stock Exchange (NSE), one of the leading stock exchanges in India. We consider the two most liquid European index options traded on the NSE, namely, NIFTY and BANKNIFTY options. The NIFTY (or Nifty 50) index consists of the 50 largest and most liquid stocks listed on the NSE, spanning diverse sectors of the Indian economy, and is calculated using the Free Float Market Capitalization weighted method. On the other hand, the BANKNIFTY (or Nifty Bank) index comprises the most liquid and large-capitalized Indian banking stocks, serving as a benchmark that reflects the capital market performance of India's banking sector. As the one month and the one week expiry options are the most liquid traded index options on the NSE, we consider the following static hedging problem for the empirical analysis. We want to hedge a target option with one month expiry using a static hedge portfolio constructed using call and put options on the same underlying expiring in one week\footnote{The python codes used for the empirical analysis are available at \\ \textit{http://github.com/Vikranth1508/FinML/tree/static\_hedge\_codes}}.

\subsection{Modelling Framework}
\label{Modelling Framework}

NSE one-month options mature on the last Thursday of a month and expire on the final Thursday of the following month. Correspondingly, one-week options are issued on all Thursdays and expire on the subsequent Thursday. If a Thursday is not a business day, the expiry is adjusted to the previous working day.

Writer of a specific one-month option (with strike $K^{*}$) aims to establish a semi-static hedge, avoiding the need for rebalancing through the week. This hedge portfolio comprises one-week options on the same underlying asset as the target option. The writer initiates the hedge portfolio every Thursday of the month and monitors daily changes in the target option value against the hedge portfolio's value throughout the week until the options in the hedge portfolio expire. This involves comparing the daily PnL of the hedge portfolio with that of the target option.

At the week's end, a new semi-static hedge is set up for the subsequent week using the proceeds from the hedge portfolio's expiring options. The performance of the hedge and target option is scrutinized weekly, except for the last week of the month, when both the target option and hedge portfolio mature simultaneously, resulting in a potential perfect hedge with an offsetting position.

The construction of the semi-static hedge involves the following steps:
\begin{enumerate}
\item {\bf Selecting candidate one-week options:} The candidate strikes for one-week call and put options in the hedge portfolio are selected based on following criteria:
\begin{itemize}
\item {\bf Trading Volume:} Selected strikes should exhibit a trade volume greater than the $50^{th}$ quantile of the trade volume of all available one-week options on the underlying index.
\item {\bf Open Interest:} Selected strikes should have an open interest greater than the $50^{th}$ quantile of open interest of all available one-week options on the underlying index.
\item {\bf Moneyness:} The candidate set considers only at-the-money (ATM) and out-of-the-money (OTM) calls and puts.
\end{itemize} 
\item {\bf Simulation:} At the commencement of each week $T_0,$ we generate $N=5000$ Monte Carlo scenarios until the weekly expiry date, $T_1.$ While various models could be contemplated for scenario generation\footnote{The decision between simulating under the risk-neutral or real-world measure is inconsequential, as our objective is to estimate $\E[V\left(S_{T_1}, T_1;K, T_2\right) \mid \bm{\phi}(S_{T_1})],$ where both $V\left(S_{T_1}, T_1;K, T_2\right)$ and $\bm{\phi}(S_{T_1})$ are $\mathcal{F}_{T_1}$ measurable.}, our study employs Equation \ref{masterprocess}, with the volatility derived from the implied volatility of the at-the-money (ATM) one-week option for clarity and consistency.
\item {\bf Pricing:} For each realisation of $S_{T_1}(\omega_i),$ where $i=1,\ldots,N,$ we compute $V\left(S_{T_1} (\omega_i), T_1;K, T_2\right).$ The pricing of this involves assuming a model, and we assess the static hedge's performance across various choices of pricing models.
\item{\bf Constructing the semi-static hedge:} With the target option values $V\left(S_{T_1} (\omega_i), T_1;K, T_2\right)$ as the dependent variable and the one-week options payoffs $\bm{\phi}(S_{T_1}(\omega_i))$ as the independent variable, where $i = 1, \ldots, N,$ the weights of the options in the hedge portfolio are determined using lasso regression. Additionally, we incorporate transaction costs for establishing long or short positions in the hedging portfolio.
\end{enumerate}

The required market data for the analysis is discussed in Section \ref{Market Data}. For simulation of index levels each week, we need the index returns and the ATM volatility of the index corresponding to one week tenor at the beginning of every week. The methodology adopted for the implied volatility surface construction is discussed in Section \ref{Implied Volatility Surface Construction}. Similarly, the model used for pricing the target option one week in the future is discussed in Section \ref{Pricing Model}.

\subsection{Pricing Model}
\label{Pricing Model}

A pricing model is essential for estimating the price of the long-term option $V\left(S_{T_1}(\omega_i), T_1;K, T_2\right)$ for the simulated scenarios $S_{T_1}(\omega_i),$ where $i = 1, \ldots, N.$ The simulated option price becomes the dependent variable, and the corresponding short maturity options payoff vector, $\bm{\phi}(S_{T_1}(\omega_i)),$ serves as the independent variable in determining the composition of the hedge portfolio through the lasso regression. We employ the Black-Scholes formula to price the target long-term option with strike $K^{*}$ maturing at $T_2$ across each simulated index level at $T_1.$ The Black-Scholes price is given by,
\begin{align}\label{BS}
V\left(S_{T_1}, T_1;K, T_2\right) &= i_{cp} \cdot \ N(i_{cp} \cdot  d_1) S_{T_1} exp\big(-q (T_2 - T_1) \big)   \\ 
& \ - \ i_{cp} \cdot N(i_{cp} \cdot d_2) K e^{-r (T_2 - T_1)}, \nonumber \\
d_1 &= \frac{ln\Big(\frac{S_{T_1}}{K} \ + \ (r - q + \frac{\sigma(S_{T_1})^{2}}{2}) (T_2 - T_1) \Big)}{\sigma(S_{T_1}) \sqrt{T_2-T_1}} , \nonumber \\
d_2 &= d_1 - \sigma(S_{T_1})\sqrt{T_2-T_1}, \nonumber
\end{align} 

 where $N(\cdot)$ is the standard normal cumulative distribution function, $q$ is the index dividend rate inferred from the corresponding index futures. The choice of modeling the volatility, $\sigma(S_{T_1})$  used in the pricing results in several model options. Through this study we also want to determine empirically if there is a benchmark model which results in a superior performance for the hedge portfolio. 

The diverse choices for modeling volatility, expressed as $\sigma(S_{T_1})$, in the pricing process introduce multiple model options. In this study, we aim to empirically identify whether there exists a benchmark model that demonstrates superior performance for the hedge portfolio.

\subsubsection{Market Data}
\label{Market Data}

The analysis covers historical data spanning one year from the last Thursday of July 2019 to the last Thursday of July 2020. This historical dataset encompasses spot values of the index, options, futures data, and information on transaction costs. These data sources are derived from contract-wise archives on the NSE platform\footnote{The market data used is accessible at \\ \textit{http://github.com/Vikranth1508/FinML/tree/static\_hedge\_mktdata}}. Risk-free interest rates are based on repo rates published by the Reserve Bank of India. The daily closing prices of the index, futures, and options are considered as the end-of-day settlement prices for their respective instruments. Specifically, futures and options data are primarily utilized for extracting settlement prices, option strikes, trading volume, and open interest.

\subsubsection{Construction of the Implied Volatility Surface}
\label{Implied Volatility Surface Construction}

In this section, we outline the methodology employed for constructing the implied volatility surface across moneyness $M$ (defined as $ M = \frac{Spot}{Strike}$) and time to maturity (refer \citet{homescu2011implied} for a comprehensive review on construction of implied volatility surface). Every Thursday $T_0$, we generate the volatility surface across the moneyness of liquid options for two tenors, $T_2 - T_0$ and $T_1 - T_0$, where $T_2$ and $T_1$, as defined in Section \ref{Static Hedging Framework}, represent the expiry dates for monthly and weekly options, respectively. For static hedging, where weights are determined only at the week's commencement, there is no need to construct the implied volatility surface on other days of the week. However, to assess the static hedge portfolio's performance against a dynamic delta hedge, we need to construct the implied volatility smile for the remainder of the tenor for the monthly option. This smile is then utilized to compute the daily delta position in the underlying, as discussed in Section \ref{Benchmark against Dynamic Delta Hedging}.

Applying the same liquidity filters as detailed in Section \ref{Modelling Framework}, we choose liquid at-the-money (ATM) and out-of-the-money (OTM) call and put options. Employing the root-finding method introduced by \citet{brent1971algorithm}, we determine the implied volatility that aligns the Black-Scholes price (Equation \ref{BS}) for an option with particular moneyness and tenor with the corresponding observed market price. The calibrated implied volatility surface is built using quotes of liquid out-of-the-money (OTM) call options for moneyness less than one. For moneyness greater than one, liquid OTM puts are used. In the at-the-money (ATM) region, the implied volatilities of both call and put options are averaged to create a smooth smile.

\begin{figure}[hbt!]
\begin{center}
\includegraphics[width=\textwidth, height=6in]{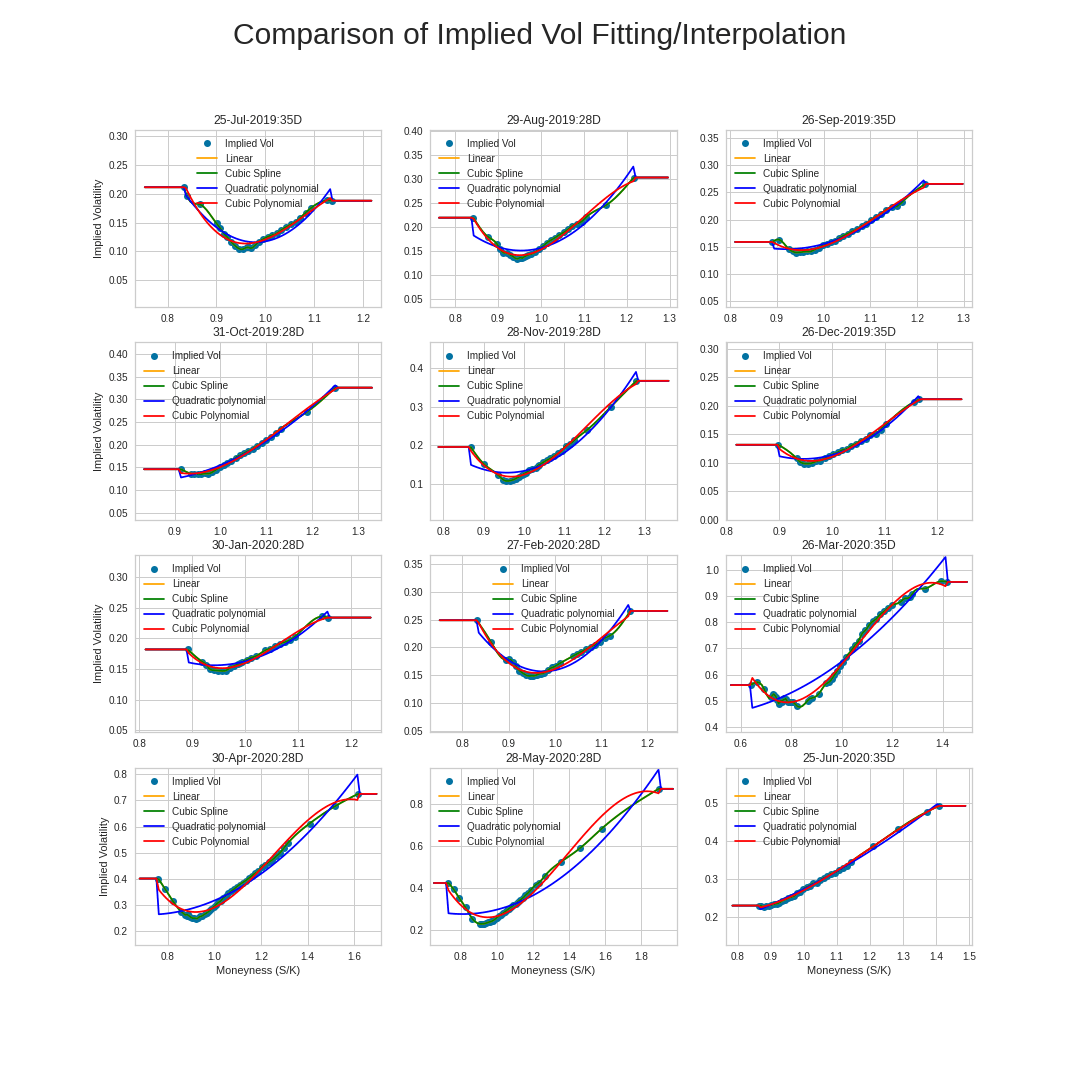}
\caption{\footnotesize Implied Volatility Smile Curve (corresponding to one-month tenor) for the NIFTY index with various interpolation fits, observed on the last Thursday of each month from July 2019 to June 2020. The x-axis represents moneyness (spot-to-strike ratio), and the y-axis represents implied volatility. Blue dots denote implied volatility inferred from liquid at-the-money (ATM) and out-of-the-money (OTM) calls and put options. Various interpolation methods—Linear, Cubic Spline, Quadratic Polynomial, and Cubic Polynomial fits—are represented by lines.} \label{A3_IV_CubicSpline_MonthlyOptions_Week0.png}
\end{center}
\end{figure}

For each time to maturity on the surface, the constructed volatility smile is anchored at specific moneyness points corresponding to the calibrated liquid options. To extend the volatility across the continuum of moneyness levels, we explore four implied volatility surface interpolation methods: 
\begin{enumerate}
\item Linear Interpolation, 
\item Cubic Spline interpolation,
\item Quadratic function fit, 
\item Cubic polynomial fit.
\end{enumerate}  

We employ an interpolation function to model the volatility smile for each tenor, and then use it to determine the volatility for a specific moneyness level associated with that tenor. Uniformly, flat extrapolation is applied across all methods.

To obtain implied volatility across the continuum of time to maturity, linear interpolation in variance-time space and flat extrapolation are employed. In Figure \ref{A3_IV_CubicSpline_MonthlyOptions_Week0.png}, a comparison of different fits of the implied volatility smile for a one-month tenor is presented on the last Thursday of each considered month. Noteworthy observations include the concentration of liquid options in normal market conditions primarily are within the moneyness range of $0.8-1.2$. However, in the Covid period, we could observe the liquidity expanding to the moneyness levels of approximately $0.6-1.4.$  Additionally, a visually better cubic fit compared to a quadratic fit is observed during the stressed period. The cubic spline and linear interpolation closely align for most regions of the smile.

%\begin{figure}[h]
%\centering
%\captionsetup{width=0.8\linewidth}
%\includegraphics[width=1.0\textwidth, height=10cm]{A3_IV_CubicSpline_MonthlyOptions_Week0.png}
%\caption{Implied Volatility Smile - Curve Fitting/Interpolation Comparison}
%\label{A3_IV_CubicSpline_MonthlyOptions_Week0.png}
%\end{figure}

\afterpage{\clearpage}

\subsubsection{Forward implied volatility}

A key model parameter, in Equation \ref{BS}, for pricing the target option on the simulated index levels at time $T_1,$ (i.e. $S_{T_1}$) is the volatility, $\sigma(S_{T_1}),$ corresponding the period $T_2 - T_1,$ given the filtration $\mathcal{F}_{T_1}.$ We consider the following four alternatives for $\sigma(S_{T_1})$:
\begin{enumerate}
\item Constant Volatility
\item Constant Volatility Smile
\item Constant Volatility Surface
\item Forward Smile
\end{enumerate}

Let $\Sigma \big(T_0; M,\tau \big)$ be the implied volatility surface constructed at $T_0$ across moneyness $M$ and  tenors, $\tau.$  The volatility surface is calibrated following the methodology detailed in Section \ref{Implied Volatility Surface Construction}.

{\bf Constant Volatility:} The constant volatility approach assumes that the volatility of the target option remains constant in the future, regardless of changes in moneyness levels. Therefore, we use $\sigma(S_{T_1}) = \Sigma(T_0; M(S_{T_0}), T_2-T_0)$, where $M(S_{T_0}) = \frac{S_{T_0}}{K^{*}}$, as the volatility for all the simulated scenarios at $T_1.$ 

{\bf Constant Smile:} In the Constant Smile approach, it is assumed that the volatility smile observed at $T_0$ for maturity $T_2$ remains unchanged over time. Consequently, the volatility for different moneyness levels realized at $T_1$ can be expressed as $\sigma(S_{T_1}(\omega_i)) = \Sigma \Big(T_0; M \big(S_{T_1}(\omega_i) \big), T_2-T_0 \Big)$, where $M \big(S_{T_1}(\omega_i) \big) = \frac{S_{T_1}(\omega_i)}{K^{*}}.$

{\bf Constant Surface:} In Constant Surface approach, it is assumed that the volatility smile for the tenor $T_2-T_1,$ remains unchanged over time. Therefore we take   $\sigma(S_{T_1}(\omega_i)) = \Sigma \Big(T_0; M \big(S_{T_1}(\omega_i) \big), T_2-T_1 \Big).$ The distinction between the constant smile and constant surface lies in the fact that, while in the former, the volatility smile corresponding to the maturity date $T_2$ remains fixed, in the latter, the volatility smile for the tenor $T_2-T_1$ remains unchanged over time.  

{\bf Forward Smile:} In the forward smile approach, the forward volatility implied based on the $T_0$ volatility surface is used to estimate $\sigma(S_{T_1}.$  The forward variance is computed as: 
$$\sigma^{2}(S_{T_1})= \Sigma^{2} \Big(T_0; M \big(S_{T_1} \big), T_2-T_0 \Big) (T_2-T_0) - \Sigma^{2} \Big(T_0; M \big(S_{T_1} \big), T_1-T_0 \Big) (T_1 - T_0) .$$

Because the moneyness level of the simulated index at $T_1$ may not align with the anchor points in the volatility surface, the volatility is determined using the interpolation schemes discussed in Section \ref{Implied Volatility Surface Construction}. The summary of the pricing model types and the corresponding volatility inputs discussed in this Section are presented in Table \ref{tab:pricing_model_options}.

% Table generated by Excel2LaTeX from sheet 'Samples'
\begin{table}[!htb]
  \centering
\resizebox{0.85\textwidth}{!}{%
    \begin{tabular}{|c|c|l|}
    \toprule
\textbf{S.No.} & \textbf{Pricing Model Type} & \textbf{Volatility Input $\sigma$} \\
    \midrule
    1        & Constant Volatility & $\Sigma(T_0; M(S_{T_0}), T_2-T_0)$, \\
	   &   & where,  $M(S_{T_0})  = \frac{S_{T_0}}{K^{*}}$ is the moneyness \\ 
& & of the target option at $T_0$. \\ 
    \midrule
    2        & Constant Volatility Smile & $\Sigma \Big(T_0; M \big(S_{T_1}(\omega_i) \big), T_2-T_0 \Big)$, \\ 
	& & where,  $M \big(S_{T_1}(\omega_i) \big) = \frac{S_{T_1}(\omega_i)}{K^{*}}$  \\ 

    \midrule
    3        & Constant Volatility Surface & $\sigma \Big(T_0; M \big(S_{T_1}(\omega_i) \big), T_2-T_1 \Big)$.  \\
    \midrule
    4        & Forward Smile Volatility & $\sigma(S_{T_1}(\omega_i))$, where, the forward variance is \\
 & &  $\sigma^{2}(S_{T_1}(\omega_i)) = \Sigma^{2} \Big(T_0; M \big(S_{T_1}(\omega_i) \big), T_2-T_0 \Big) (T_2-T_0)$ \\ 
 & & $ \ \ \ \ \ \ \ \ \  - \Sigma^{2} \Big(T_0; M \big(S_{T_1}(\omega_i) \big), T_1-T_0 \Big) (T_1 - T_0)  $. \\
    \bottomrule
    \end{tabular}%
}
  \caption{\footnotesize Target Options Pricing Model Types}
  \label{tab:pricing_model_options}%
\end{table}%

\subsection{Model Alternatives and Test Scenarios}
\label{Model Alternatives and Test Scenarios}

To construct the static hedge portfolio of short maturity options expiring at $T_1,$ we must determine the target option value $V\left(S_{T_1}(\omega_i), T_1;K, T_2\right)$ for the simulated index level $S_{T_1}(\omega_i)$ at $T_1$ across scenarios $i = 1, \ldots, N.$ To utilize the Black-Scholes model for pricing the option on the future date, the following would be required
\begin{enumerate}
\item A model for future volatility $\sigma(S_{T_1}(\omega_i)).$
\item An interpolation scheme for the above to obtain volatility at any arbitrary simulated moneyness level $M(S_{T_1}(\omega_i)).$
\end{enumerate}

The Table \ref{tab:model_options} enlists the considered alternative for the above two model choices. A specific implementation for the construction of the static hedge would be a combination of the two choices, for instance using (a) constant volatility smile with volatility values interpolated using the (b) cubic spline method for a simulated moneyness level.  There are therefore sixteen variations that leads to sixteen different static hedging model choices.

% Table generated by Excel2LaTeX from sheet 'Samples'
\begin{table}[!htb]
  \centering
\resizebox{0.85\textwidth}{!}{%
    \begin{tabular}{|c|c|c|}
    \toprule
   \textbf{S.No.} & $\sigma(S_{T_1}(\omega_i))$ & \textbf{Interpolation scheme} \\
    \midrule
    1        & Constant Volatility & Linear Interpolation \\
    \midrule
    2        & Constant Volatility Smile & Cubic Spline \\
    \midrule
    3        & Constant Volatility Surface & Quadratic Function \\
    \midrule
    4        & Forward Smile Volatility & Cubic Polynomial  \\
    \bottomrule
    \end{tabular}%
}
  \caption{\footnotesize Static Hedge Model Options}
  \label{tab:model_options}%
\end{table}%

The analysis is conducted for the following cases: call and put options written on BANKNIFTY and NIFTY indices. For each option type, such as a put option on BANKNIFTY, three potential scenarios are considered, namely Out-of-the-Money (OTM), At-the-Money (ATM), and In-the-Money (ITM), representing the initial moneyness levels of the monthly option subjected to hedging. The option with moneyness $M(S_{T_0})$ nearest to 1 is designated as ATM, while those at 0.9 and 1.1 are categorized as OTM and ITM, respectively, for call options. Correspondingly, a similar categorization is applied to put options. To illustrate, in the instance of an ATM BANKNIFTY call option, we select the option with one month expiry that is ATM at the commencement of the month. We monitor the daily performance of the hedge until the option's maturity. A semi-static hedge is established at the conclusion of each week, utilizing the available weekly options. Subsequently, at the beginning of the next month, we begin with a new ATM BANKNIFTY call option. The evaluation of hedge performance is undertaken in two distinct market scenarios: one encompassing the entire historical dataset and the other confined to the Covid historical window. The Covid window is chosen to study the hedge performance in a period marked by jumps and high volatility. 

For each case, we compare the performance of the different variants of the Lasso static hedge, dynamic hedging, and Carr Wu static hedge.

\subsection{The test for Superior Predictive Ability}
\label{Superior Predictive Analysis}

Comparing outcomes from multiple models gives rise to a non-standard inference challenge. Unless the inference appropriately adjusts for multiple comparisons, there is a likelihood of encountering spurious results. The likelihood of an inferior model outperforming others by chance increases with the number of models under comparison. To address whether a particular benchmark is significantly outperformed by any of the alternatives used in the comparison, we use the test for superior predictive ability of \cite{hansen2005test}. The test assesses whether a specific model, acting as a benchmark, is significantly surpassed by other models, considering the extensive set of compared models. Essentially, the test determines the significance of observed excess performance, distinguishing it from random occurrences. 

 Initially, we employ the SPA test to distinguish the superior models within the variations of both (a) the LASSO static hedge and (b) the Dynamic delta hedge individually. Subsequently, we compare the superior models of LASSO static hedge, dynamic hedge, and Carr-Wu static hedge. These comparisons are carried out individually for each of the following: ITM, ATM, and OTM call and put options written on the NIFTY and BANKNIFTY indices. The evaluation spans the entire period from July 2019 to July 2020, and separately the Covid period from February 27, 2020, to July 20, 2020. Therefore in total there are $ 3 \times 2 \times 2 \times 2 = 24$ universes in which the models are tested for superiority. 
 
 The SPA test entails comparing the historical loss time series of the benchmark model with that of the alternative models.  Loss is defined in terms of the difference in the daily changes in the market value of the target option and the corresponding changes in the market value of the  hedge portfolio. When the hedge effectively replicates the target option, this difference tends to zero. 

 {\bf PnL time series:} The daily PnL of a portfolio is the difference between its closing market value on a given day and its previous closing price.  It's important to note that there are no model assumptions involved in computing the PnLs; rather, they are directly derived from the closing prices of market instruments, as explained in the following points.  Figure \ref{SPAtimelines.png}  illustrates the generation of value time series for the target option and static hedge portfolio over a one-week period. Subsequently, the daily PnL time series is calculated based on the corresponding value time series. For each time $t$, the target and hedge PnL are represented as:

\begin{figure}[hbt!]
\begin{center}
\includegraphics[width=5in, height=2.9in]{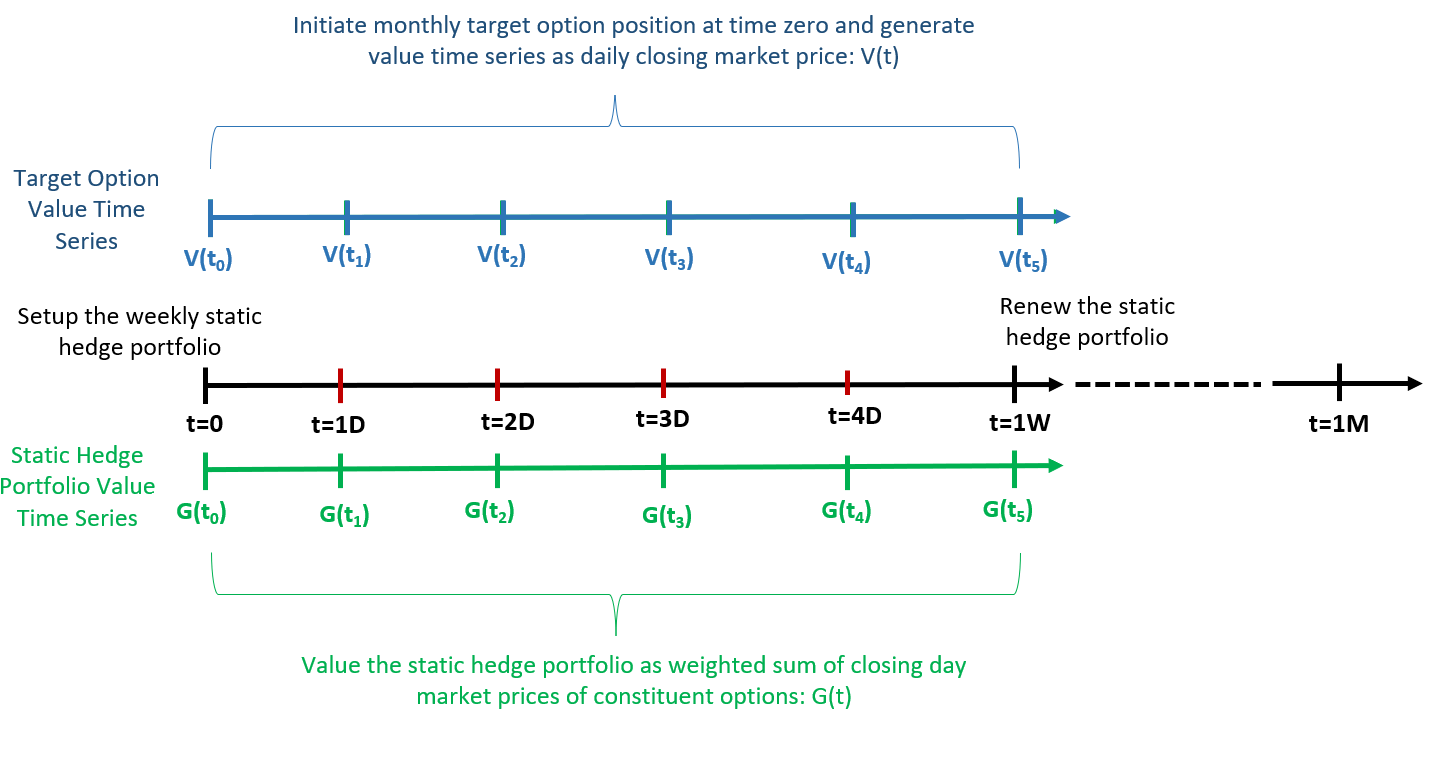}
\caption{\footnotesize The figure depicts a one-week snapshot of the value time series generation for the monthly target option $V(t_k)$ and the weekly static hedge portfolio $G(t_k)$. The black line represents the time axis, the green line illustrates the value time series of the static hedge portfolio, and the blue line represents the value of the target option. Both the static hedge portfolio and target option positions are initiated at time zero. The target option value over time corresponds to the closing day market price, while the static hedge portfolio value is the weighted sum of the closing day market prices of its constituent options. At the conclusion of each week, the static hedge portfolio is refreshed, and the value time series is generated using the same logic.} \label{SPAtimelines.png}
\end{center}
\end{figure}

\begin{enumerate}

\item {\bf The targe option PnL:} $\Delta V_{t}$ \\
The target option PnL is the difference of daily realised closing price of the position in target option entered at time zero ($t_0$). Define at  time $t_k$, $\Delta V_{t_k} = V_{t_k} - V_{t_{k-1}} $, where, $V_{t_k}$ and $V_{t_{k-1}}$ are the closing market price of the target option as of time $t_k$ and $t_{k-1}$ respectively. Therefore, the PnL generation is completely model independent and evaluated from realised end-of-day market price.

\item {\bf The hedge portfolio PnL:} $\Delta \mathcal{G}_{m, t}$, where, $m = 0, 1, \ldots, M-1,$ and $M$ is the total number of models to compare.

\begin{itemize}

\item {\bf Static hedge portfolio PnL:} For a given model under evaluation, the static hedge is constructed at time $t_0.$  The market value of the hedge portfolio is determined as the weighted sum of the market value of its individual short-term options, where the weights are obtained from the model at $t_0.$ The daily portfolio value is computed for the week using the same portfolio weights and constituent options as set up at $t_0$. At the end of each week, the composition of the static hedge portfolio is refreshed.  The daily PnL for the static hedge portfolio is the difference of portfolio value between two consecutive days ($\Delta \mathcal{G}_{m, t_k} = \mathcal{G}_{m, t_k} - \mathcal{G}_{m, t_{k-1}}$). Once again, the hedge PnL time series is model-independent and constructed using the realized end-of-day market prices of the constituent short-term options.

\item {\bf Dynamic hedge portfolio PnL:} The dynamic hedge portfolio is set up every day by taking the target option delta (partial derivative of target option value with respect to the underlying) number of positions in the underlying and a money market account (or bank account) to match the target option value. Everyday, the portfolio is refreshed based on the option delta as of that day. The delta hedge PnL is therefore change in value of the delta hedge portfolio set up on the previous day i.e., sum of the change in closing price of the underlying postions entered and the change in money market account value. The PnL is driven by the movement of the realised  underlying index's closing market price and the constant interest income or expenditure in the money market account.

\end{itemize}

\end{enumerate}

 {\bf Loss Function:} Each modelling approach yields the loss sequence, $L_{m,t} = L(\Delta \mathcal{G}_{m, t}, \Delta V_{t})$. We consider two different loss functions $L(.)$:

\begin{itemize}
 \item Absolute Error Loss Function:  $L_{m,t} = || \Delta \mathcal{G}_{m, t} - \Delta V_{t}||_{1}$,
 \item Squared  Error Loss unction. $L_{m,t} = || \Delta \mathcal{G}_{m, t} - \Delta V_{t}||_{2}^{2}$.
\end{itemize}

The squared loss function computes the sequence of squared differences, while the absolute error loss function calculates the sequence of absolute differences between the PnL of the hedge model and the PnL of the target option. Therefore, the loss basically measure how change in the hedge portfolio offsets change in target option value, and for a perfect hedge, it would be close to zero. Further, for setting up the static hedge, we used the information at beginning of the week. Once the weights are obtained for the static hedge portfolio, we can compute  the out of sample loss for the corresponding week using only the closing prices. 

For the SPA test, the first model $m=0$ is considered the benchmark model that is compared to the other alternative models. We define the relative performance variables as: 

      $R_{m,t} = (L_{0,t} - L_{m,t})$,  $m = 1, 2, \ldots, M$.

 {\bf Null Hypothesis:} $H_0 : \boldsymbol{\mu_m}  \triangleq \mathbb{E}[R_{m, t}]) \le 0 $ for all $m = 1, 2, \ldots, M$. In other words, the null hypothesis is that the benchmark model is not inferior to any of the alternative models. 

%This test uses a studentised  test statistic, whose sampling distribution is estimated using stationary bootstrapping developed by \citet{politis1994stationary} with optimal block length defined by \citet{politis2004automatic} with the support from \citet{gonccalves2003consistency}, which shows the convergence of empirical distribution to the true asymptotic distribution. This methodology requires stationarity of the performance sequence and empirically tested the stationarity of the performance series using Augmented Dickey-Fuller (ADF) test and Kwiatkowski–Phillips–Schmidt– Shin (KPSS) test under 95\% confidence. Across all model comparisons in Sections \ref{Selection of Superior Lasso Static hedge Model}-\ref{Benchmark against Carr-Wu Static hedge}, we observe that performance time series is stationary for squared error loss function and there are a very few exceptions for absolute loss function among all the comparisons - four exceptions within 1440 lasso static hedging comparisons and four exceptions in NIFTY ITM Put Options within the dynamic hedging comparisons. The hypothesis test is performed at 95\% confidence level, in other words, the null hypothesis is rejected if p-value is less than 5\%. The p-values consistent to true p-value is used along with lower and upper boud p-values. For more theoritcal explanation and assumptions of SPA test, refer the work by \citet{hansen2005test}. \\

 This test employs a studentized test statistic, the sampling distribution of which is estimated through stationary bootstrapping as developed by \citet{politis1994stationary}, with optimal block length defined by \citet{politis2004automatic}  and supported by \citet{gonccalves2003consistency}, demonstrating the convergence of the empirical distribution to the true asymptotic distribution. This methodology necessitates the stationarity of the performance sequence, empirically confirmed in this work through Augmented Dickey-Fuller (ADF) test and Kwiatkowski–Phillips–Schmidt– Shin (KPSS) test under a 95\% confidence level. Across all model comparisons in Sections \ref{Selection of Superior Lasso Static hedge Model}-\ref{Benchmark against Carr-Wu Static hedge}, we observe that the performance time series is stationary for the squared error loss function, with very few exceptions for the absolute loss function. Specifically, there are four exceptions within 1440 lasso static hedging comparisons and four exceptions in NIFTY ITM Put Options within the dynamic hedging comparisons. 

 The hypothesis test is conducted at a 95\% confidence level, meaning that the null hypothesis is rejected if the p-value is less than 5\%. The p-values consistent with the true p-value are used, along with lower and upper bound p-values. For a more in-depth theoretical explanation and assumptions of the SPA test, please refer to the work by \citet{hansen2005test}. 

 {\bf Best Model Selection Criteria:} We establish here the criteria for identifying the best or superior model from the pool of models under consideration. We perform SPA with each model approach as a benchmark and compare it against other models. We consider a model is universally superior among the set of models if two conditions are satisfied:

\begin{itemize}
 \item If we cannot reject the null hypothesis of the SPA (Consistent) test with the corresponding model as the benchmark model (and others as alternatives) under all the universes (based on different underlying, option types, moneyness, and error loss functions). 
\item If all other alternative models have at least one rejection case of the null hypothesis when used as a SPA (Consistent) test benchmark. 
\end{itemize}

\subsubsection{Selection of Superior Lasso Static hedge Models}
\label{Selection of Superior Lasso Static hedge Model}

%We performed SPA with each one of the 16 lasso-based static hedge models as a benchmark and compared it against other models. We consider a model is universally superior among a set of models if two conditions are satisfied:
%\begin{itemize}
% \item If we cannot reject the null hypothesis of the SPA (Consistent) test with the corresponding model as the benchmark model (and others as alternatives) under all the testing scenarios (based on different underlying, option types, moneyness, and error loss functions). 
%\item If all other alternative models have at least one rejection case of the null hypothesis when used as a SPA (Consistent) test benchmark. 
%\end{itemize}

% Table generated by Excel2LaTeX from sheet 'Results Tables'
\begin{table}[htb]
  \centering
\resizebox{0.85\textwidth}{!}{%
    \begin{tabular}{|c|p{3.3em}|p{3em}|c|c|c|c|c|c|}
    \toprule
    \multicolumn{9}{|c|}{\textbf{Static Hedging Comparisons P-Values}} \\
    \midrule
    \multicolumn{9}{|c|}{\textbf{Benchmark: Constant Volatility Linear Model}} \\
    \midrule
    \multicolumn{1}{|c|}{\multirow{3}[4]{*}{\textbf{Index}}} & {\multirow{3}[4]{*}{\parbox{1cm}{\textbf{Option \\ Type  }}}} & \multirow{3}[4]{*}{\parbox{1cm}{\textbf{Money \\ ness}}} & \multicolumn{3}{c|}{\textbf{Absolute Error Loss Function}} & \multicolumn{3}{c|}{\textbf{Squared Error Loss Function}} \\
\cmidrule{4-9}             &  &         & \textbf{SPA } & \textbf{SPA } & \textbf{SPA } & \textbf{SPA } & \textbf{SPA } & \textbf{SPA } \\
\cmidrule{4-9}             &  &        & \textbf{Lower} & \textbf{Consistent} & \textbf{ Upper} & \textbf{Lower} & \textbf{Consistent} & \textbf{Upper} \\
    \midrule
    \multicolumn{1}{|c|}{\multirow{6}[12]{*}{NIFTY}} & \multicolumn{1}{c|}{\multirow{3}[6]{*}{Call}} & ATM      & 0.63     & 0.93     & 0.95     & 0.64     & 0.94     & 0.97 \\
\cmidrule{3-9}             &          & ITM      & 0.38     & 0.71     & 0.71     & 0.25     & 0.69     & 0.69 \\
\cmidrule{3-9}             &          & OTM      & 0.70     & 0.97     & 0.97     & 0.53     & 0.93     & 0.93 \\
\cmidrule{2-9}             & \multicolumn{1}{c|}{\multirow{3}[6]{*}{Put}} & ATM      & 0.63     & 0.77     & 0.95     & 0.66     & 0.95     & 0.96 \\
\cmidrule{3-9}             &          & ITM      & 0.74     & 0.93     & 0.93     & 0.61     & 0.94     & 0.94 \\
\cmidrule{3-9}             &          & OTM      & 0.19     & 0.43     & 0.59     & 0.24     & 0.46     & 0.46 \\
    \midrule
    \multicolumn{1}{|c|}{\multirow{6}[12]{*}{BANKNIFTY}} & \multicolumn{1}{c|}{\multirow{3}[6]{*}{Call}} & ATM      & 0.71     & 0.84     & 1.00     & 0.67     & 0.99     & 0.99 \\
\cmidrule{3-9}             &          & ITM      & 0.49     & 0.76     & 0.90     & 0.38     & 0.72     & 0.72 \\
\cmidrule{3-9}             &          & OTM      & 0.76     & 0.88     & 0.96     & 0.56     & 0.91     & 0.91 \\
\cmidrule{2-9}             & \multicolumn{1}{c|}{\multirow{3}[6]{*}{Put}} & ATM      & 0.66     & 0.98     & 1.00     & 0.56     & 0.99     & 0.99 \\
\cmidrule{3-9}             &          & ITM      & 0.73     & 0.99     & 1.00     & 0.58     & 0.83     & 0.95 \\
\cmidrule{3-9}             &          & OTM      & 0.10     & 0.41     & 0.65     & 0.13     & 0.63     & 0.63 \\
    \bottomrule
    \end{tabular}%
}
  \caption{\footnotesize SPA Comparison among Static Hedge Model alternatives based on forward volatility models and implied volatility surface interpolation schemes. The benchmark model is Constant Volatility Linear model. The test is conducted with NIFTY and BANKNIFTY call and put options across ATM, ITM and OTM moneyness levels employing two loss functions: Absolute and Squared Erros loss function. In the table, each value corresponds to the p-value associated with a specific combination of option characteristics and loss function. The test is performed at 95\% confidence level.}
  \label{tab:static_winner}%
\end{table}%

 Among the sixteen variants of Lasso-based static hedge models, as detailed in Section \ref{Model Alternatives and Test Scenarios}, our observation indicates that the Constant Volatility Linear Model\footnote{The Constant Volatility Linear Model is a Lasso static hedge constructed using \emph{constant volatility} as the input for pricing the target option in simulated scenarios. The volatility is extracted from the calibrated volatility surface through \emph{linear interpolation}.} stands out as the universally superior model across all 24 universes. SPA P-values (Lower, Consistent, Upper) for the benchmark Lasso Constant Volatility Linear Model with both Absolute Error Loss and Squared Error Loss functions are presented in Table \ref{tab:static_winner}. In contrast, when considering the other static hedge models as the benchmark, we observed the null hypothesis being rejected in  at least one of the universes. 

\begin{figure}[hbt!]
\begin{center}
\includegraphics[width=\textwidth, height=5in]{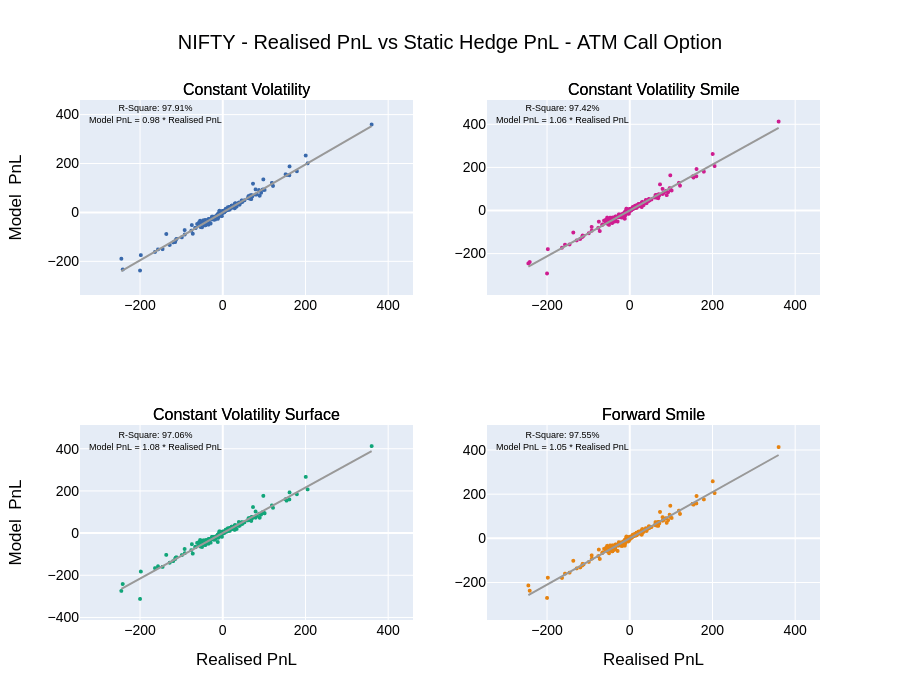}
\caption{\footnotesize The figure illustrates the regression performance of the static hedging Profit and Loss (PnL) in comparison to the target PnL of NIFTY ATM CALL option. In all subplots, the x-axis represents the realized PnL of the target option, while the y-axis depicts the realized PnL of the static hedge model portfolio over a one-year historical period. Each subplot corresponds to a specific static hedge model based on the forward volatility used in pricing, and all four subplots align with the Linear interpolation scheme of the implied volatility surface.} \label{P2_NIFTY_Linear_ATM_CE_static_regression_fit.png}
\end{center}
\end{figure}

 We show for representative purpose, the regression fit between target option and static hedge realised PnLs obtained by four different types of forward implied volatility models (with linearly interpolated calibrated volatility surface) for the NIFTY ATM Call option in Figure  \ref{P2_NIFTY_Linear_ATM_CE_static_regression_fit.png}. We observe that target option PnL and constant volatility static hedge model PnL closely align. We see a good alignment for other alternative models, but the R-Square is slightly lower and the beta coefficient is slightly farther from $1$ than the constant volatility hedge model. We also observed similar behaviour of constant volatility static hedge in NIFTY and BANKNIFTY call and put options for different moneyness levels.
%Further, in Section \ref{Benchmark against Dynamic Delta Hedging}, we compare static hedging errors with dynamic hedge errors.

%\begin{figure}[h]
%\centering
%\captionsetup{width=0.9\linewidth}
%%\includegraphics[width=0.8\textwidth, height=0.5\textheight]
%\includegraphics[width=0.9\textwidth, height=9cm]
%{P2_BANKNIFTY_Linear_ATM_CE_static_regression_fit.png}
%\caption{BANKNIFTY ATM CALL Option: Static Hedging PnL Regression Fits}
%\label{P2_BANKNIFTY_Linear_ATM_CE_static_regression_fit.png}
%\end{figure}

%\begin{figure}[h]
%\centering
%\captionsetup{width=0.9\linewidth}
%\includegraphics[width=0.9\textwidth, height=9cm]{P2_NIFTY_Linear_ATM_CE_static_regression_fit.png}
%\caption{NIFTY ATM CALL Option: Static Hedging PnL Regression Fits}
%\label{P2_NIFTY_Linear_ATM_CE_static_regression_fit.png}
%\end{figure}

\subsubsection{Outperformance in Hedging: Lasso Static Hedge versus Dynamic Delta Hedge}
\label{Benchmark against Dynamic Delta Hedging}

\begin{figure}[!htb]
\begin{center}
\includegraphics[width=\textwidth, height=3in]{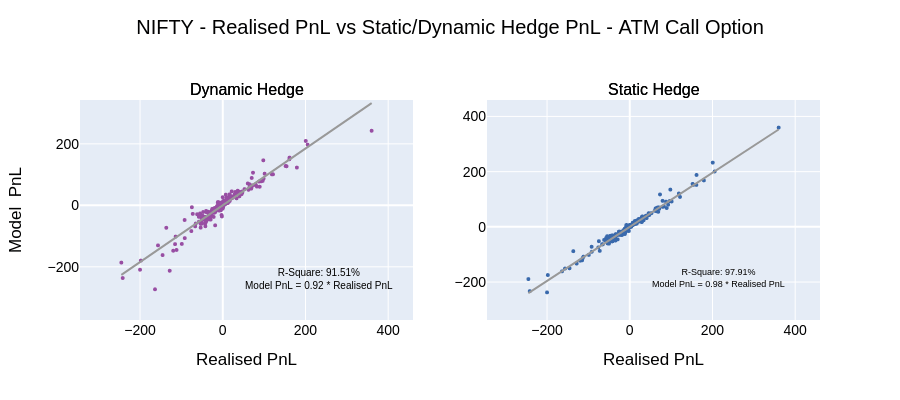}
\caption{\footnotesize The figure compares the regression performance of static hedging Profit and Loss (PnL) and dynamic hedge PnL with the target PnL of NIFTY ATM CALL option. In each subplot, the x-axis reflects the realized PnL of the target option, while the y-axis shows the realized PnL of the hedge model portfolio over a one-year historical period. The first subplot focuses on the dynamic hedge model PnL compared to the target option PnL, while the second subplot assesses the performance of the constant linear static hedge model against the target option PnL.} \label{P2_NIFTY_Linear_ATM_CE_dynamic_regression_fit.png}
\end{center}
\end{figure}

\begin{figure}[hbt!]
\begin{center}
\includegraphics[width=\textwidth, height=5in]{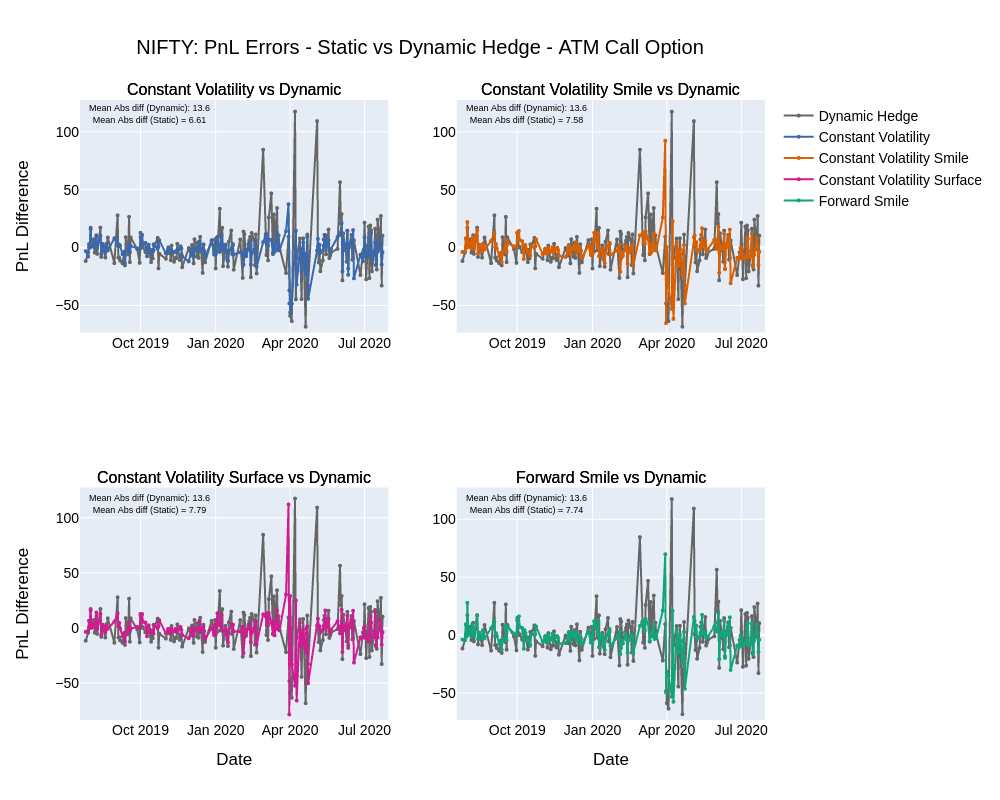}
\caption{\footnotesize The figure compares the static hedging Profit and Loss (PnL) error and dynamic hedging PnL error time series for NIFTY ATM CALL option. The PnL error of the hedge model is calculated as the difference between realised target option PnL and realised hedge model portfolio PnL. In all subplots, the x-axis represents the historical dates, while the y-axis depicts the PnL error of the hedge model portfolio over a one-year historical period. Each subplot corresponds to a specific static hedge model based on the forward volatility used in pricing, and all four subplots align with the Linear interpolation scheme of the implied volatility surface. In each subplot, a specific static hedge model alternative is compared against the dynamic hedge model.} \label{P2_NIFTY_Linear_ATM_CE_static_vs_dynamic_pnL_error.png}
\end{center}
\end{figure}

In this section, we first look at the different dynamic hedging model alternatives based on the interpolation methodology used to obtain the implied volatility to calculate the delta. Then, we compare the winners within static hedging and dynamic hedging model universe.  

 To perform dynamic hedging, we assume that the underlying index is traded and we set up the dynamic hedge portfolio daily by matching the target option value with delta positions in the underlying and the remaining in a deposit or money market account (with the risk-free interest rate for borrowing and lending). To compute the delta of the target option, we require the corresponding target option volatility based on its current moneyness level. This volatility is obtained from the daily constructed volatility surface using the four interpolation schemes discussed earlier. Initially, we identify the superior dynamic hedge model among the four alternatives, and it turns out to be the Linear dynamic hedge model\footnote{In the Linear dynamic hedge model, the volatility for delta calculation is obtained through linear interpolation of the calibrated implied volatility surface} (Refer to Appendix \ref{App:Benchmark of Lasso Static Hedge with Dynamic delta hedge}, Table \ref{tab:dynamic_winner} for SPA P-values).

 After determining the winning model from the dynamic hedging universe, we conduct a SPA comparison between the Constant Volatility Linear Static Hedge model and the selected dynamic hedging model. Our observation reveals that the Constant Volatility Linear static hedge consistently outperforms under both full market and Covid market conditions. The SPA results with the static hedge as the benchmark for the entire one-year historical window are provided in Appendix \ref{App:Benchmark of Lasso Static Hedge with Dynamic delta hedge}, Table \ref{tab:stat_dyn_winner}. Although not explicitly reported, the same outcome is observed for the Covid period. In the case of NIFTY ATM Call options, as depicted in Figure \ref{P2_NIFTY_Linear_ATM_CE_dynamic_regression_fit.png}, it is clear that the static hedge PnL exhibits a superior fit with the Realised PnL, as indicated by the R-squared values, in comparison to the dynamic hedge PnL. Similar outcomes are observed for all the other universes (combination of underlying, option type, and moneyness). 

Figure \ref{P2_NIFTY_Linear_ATM_CE_static_vs_dynamic_pnL_error.png} compares the PnL errors of various Lasso static hedge models with those of the dynamic hedge, specifically for the NIFTY ATM case. The error is calculated as the difference between the realized target option PnL and the realized hedge model portfolio PnL. It is evident that the static hedge outperforms the dynamic hedge, with spikes observed in dynamic hedge errors, particularly during the Covid period. The spikes are notably subdued in the case of the Lasso static hedge. This consistent trend in static hedge models is observed across other option types, underlying indices, and moneyness levels.

\subsubsection{Benchmark against Carr-Wu Static hedge}
\label{Benchmark against Carr-Wu Static hedge}

% Table generated by Excel2LaTeX from sheet 'winner-call'
\begin{table}[hbt!]
  \centering
\resizebox{0.85\textwidth}{!}{%
    \begin{tabular}{|c|p{3em}|c|c|c|c|c|c|}
    \toprule
    \multicolumn{8}{|c|}{\textbf{SPA (P-Values) : Lasso Static Hedge , Dynamic Hedge, Carr Static Hedge (Call Options)}} \\
    \midrule
    \multicolumn{8}{|c|}{\textbf{Benchmark: Constant Volatility Linear Static Hedge Model}} \\
    \midrule

    \multicolumn{1}{|c|}{\multirow{3}[4]{*}{\textbf{Index}}}  & \multirow{3}[4]{*}{\parbox{1cm}{\textbf{Money \\ ness}}} & \multicolumn{3}{c|}{\textbf{Absolute Error Loss Function}} & \multicolumn{3}{c|}{\textbf{Squared Error Loss Function}} \\
\cmidrule{3-8}               &         & \textbf{SPA } & \textbf{SPA } & \textbf{SPA } & \textbf{SPA } & \textbf{SPA } & \textbf{SPA } \\
\cmidrule{3-8}               &        & \textbf{Lower} & \textbf{Consistent} & \textbf{ Upper} & \textbf{Lower} & \textbf{Consistent} & \textbf{Upper} \\

    \midrule
   \multicolumn{1}{|c|}{\multirow{3}[6]{*}{NIFTY}} & ATM      & 0.51     & 0.51     & 1.00     & 0.51     & 0.51     & 1.00 \\
\cmidrule{2-8}             & ITM      & 0.52     & 0.52     & 1.00     & 0.56     & 0.56     & 1.00 \\
\cmidrule{2-8}             & OTM      & 0.55     & 0.89     & 0.93     & 0.65     & 0.93     & 0.93 \\
    \midrule
    \multicolumn{1}{|c|}{\multirow{3}[6]{*}{BANKNIFTY}} & ATM      & 0.63     & 0.63     & 1.00     & 0.75     & 0.75     & 1.00 \\
\cmidrule{2-8}             & ITM      & 0.52     & 0.52     & 1.00     & 0.54     & 0.54     & 1.00 \\
\cmidrule{2-8}             & OTM      & 0.53     & 0.94     & 0.99     & 0.49     & 0.78     & 0.92 \\
    \bottomrule
    \end{tabular}%
}
  \caption{\footnotesize SPA comparison among Constant Linear Static Hedge Model, Dynamic Hedge Model and Carr-Wu static hedge model with linear interpolation employed for implied volatility surface construction in all the models. The benchmark model is Constant Volatility Linear static hedge model. The test is conducted with NIFTY and BANKNIFTY call options across ATM, ITM and OTM moneyness levels employing two loss functions: Absolute and Squared Erros loss function. In the table, each value corresponds to the p-value associated with a specific combination of option characteristics and loss function. The test is performed at 95\% confidence level.}
  \label{tab:stat_dyn_carr_winner}%
\end{table}%

\begin{figure}[hbt!]
\begin{center}
\includegraphics[width=\textwidth, height=3in]{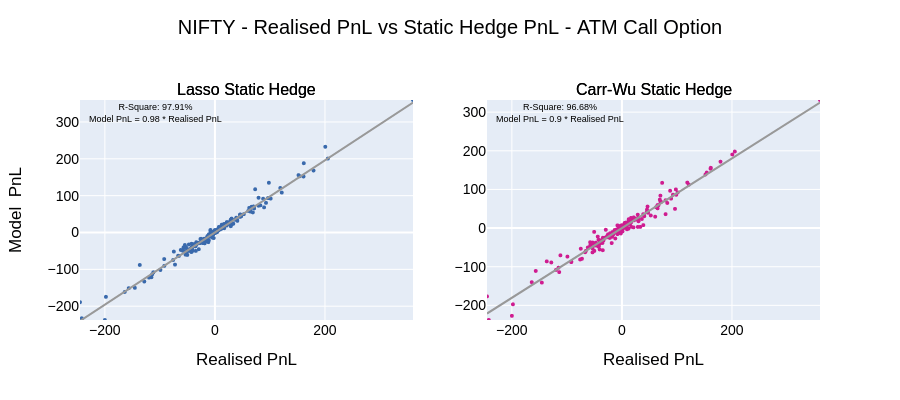}
\caption{\footnotesize The figure compares the regression performance of Lasso static hedge Profit and Loss (PnL) and Carr-Wu static hedge PnL with the target PnL of NIFTY ATM CALL option. In each subplot, the x-axis reflects the realized PnL of the target option, while the y-axis shows the realized PnL of the hedge model portfolio over a one-year historical period. The first subplot focuses on the Carr-Wu static hedge model PnL compared to the target option PnL, while the second subplot assesses the performance of the constant linear Lasso static hedge model against the target option PnL.} \label{P2_NIFTY_Linear_ATM_CE_static_regression_fit_carrWu.png}
\end{center}
\end{figure}

\begin{figure}[hbt!]
\begin{center}
\includegraphics[width=\textwidth, height=3in]{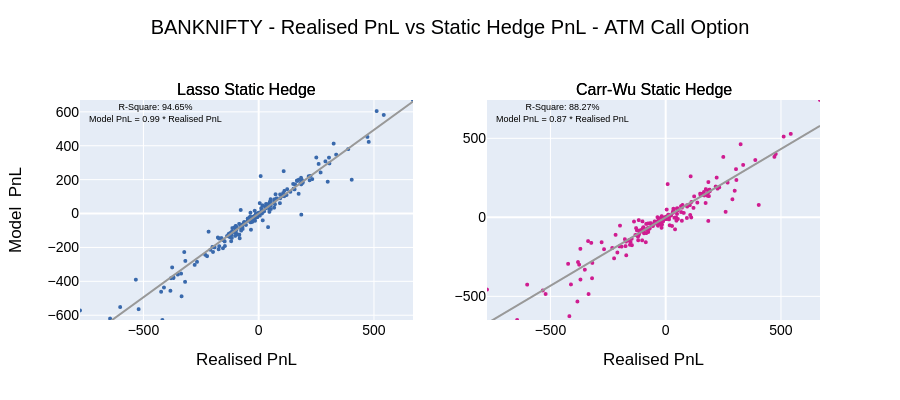}
\caption{\footnotesize The figure compares the regression performance of Lasso static hedge Profit and Loss (PnL) and Carr-Wu static hedge PnL with the target PnL of BANKNIFTY ATM CALL option. In each subplot, the x-axis reflects the realized PnL of the target option, while the y-axis shows the realized PnL of the hedge model portfolio over a one-year historical period. The first subplot focuses on the Carr-Wu static hedge model PnL compared to the target option PnL, while the second subplot assesses the performance of the constant linear Lasso static hedge model against the target option PnL.} \label{P2_BANKNIFTY_Linear_ATM_CE_static_regression_fit_carrWu.png}
\end{center}
\end{figure}

We construct the Carr-Wu static hedge portfolio (with 10 constituent shorter-term call options) for the target longer-term call options by generating the portfolio weights and the theoretical strikes of shorter-term options by using the Gauss-Hermite quadrature as proposed in \citet{carr2014static}. Once the theoretical strikes of the options are obtained, we match them to the closest available strikes of liquid options in the market and use the corresponding portfolio weights. 

We conduct the SPA test among Lasso constant volatility static hedge, dynamic hedge, and the Carr-Wu static hedge—all models utilizing linear interpolation on the calibrated implied volatility surface. Our observations reveal that the Lasso static hedge consistently outperforms the other models. The P-values of the SPA test with the Lasso static hedge as the benchmark are presented in Table \ref{tab:stat_dyn_carr_winner}. Additionally, we compare the regression fits of Lasso and Carr-Wu static hedge PnL against realized PnL for NIFTY and BANKNIFTY ATM Call options in Figures \ref{P2_NIFTY_Linear_ATM_CE_static_regression_fit_carrWu.png} and \ref{P2_BANKNIFTY_Linear_ATM_CE_static_regression_fit_carrWu.png}. While the Carr-Wu static hedge PnL fits well for NIFTY options, its performance is slightly inferior for BANKNIFTY options. In both cases, the fit with the Lasso static hedge is relatively better. This observation might be attributed to the fact that the Carr-Wu static hedge relies on the availability of liquid option strikes close to theoretical strikes, which is not always guaranteed.

%\begin{figure}[h]
%\centering
%\captionsetup{width=0.8\linewidth}
%\includegraphics[width=0.8\textwidth, height=5cm]{P2_NIFTY_Linear_ATM_CE_static_regression_fit_carrWu.png}
%\caption{BANKNIFTY ATM CALL Option: Lasso vs Carr Wu Static Hedging PnL Regression Fits}
%\label{P2_NIFTY_Linear_ATM_CE_static_regression_fit_carrWu.png}
%\end{figure}

%\begin{figure}[h]
%\centering
%\captionsetup{width=0.8\linewidth}
%\includegraphics[width=0.8\textwidth, height=5cm]{P2_BANKNIFTY_Linear_ATM_CE_static_regression_fit_carrWu.png}
%\caption{BANKNIFTY ATM CALL Option: Lasso vs Carr Wu Static Hedging PnL Regression Fits}
%\label{P2_BANKNIFTY_Linear_ATM_CE_static_regression_fit_carrWu.png}
%\end{figure}

%\begin{figure}[h]
%\centering
%\captionsetup{width=0.8\linewidth}
%\includegraphics[width=0.8\textwidth, height=5cm]{P2_BANKNIFTY_Linear_ATM_CE_carrWu_lasso_pnL_error.png}
%\caption{BANKNIFTY ATM CALL Option: Lasso vs Carr Wu Static Hedging Hedging PnL Errors}
%\label{P2_BANKNIFTY_Linear_ATM_CE_carrWu_lasso_pnL_error.png}
%\end{figure}

%\begin{figure}[h]
%\centering
%\captionsetup{width=0.8\linewidth}
%\includegraphics[width=0.8\textwidth, height=5cm]{P2_NIFTY_Linear_ATM_CE_carrWu_lasso_pnL_error.png}
%\caption{NIFTY ATM CALL Option: Lasso vs Carr Wu Static Hedging Hedging PnL Errors}
%\label{P2_NIFTY_Linear_ATM_CE_carrWu_lasso_pnL_error.png}
%\end{figure}

\section{Analysis on Static Hedge Performance}

In this section, we conduct PnL attribution to identify the factors contributing to the outstanding performance of static hedging, as highlighted by the SPA tests and PnL errors presented in Section \ref{Empirical Analysis of Lasso Static Hedge}. Of particular interest is the comparative performance of static hedge and dynamic hedge during the Covid period.

We begin by establishing the framework for the PnL attribution along with the corresponding notations. For the purpose of this analysis, we assume a constant risk-free interest rate and zero dividends. The concept of three distinct PnLs --full PnL, risk PnL, and marginal PnL-- is discussed below.

\begin{enumerate}

\item The full PnL refers to the difference between the realised closing price or realised value of a specific portfolio or instrument of interest on the current business day ($t$) and the previous business day ($t-1$). Let realised closing price of the target option at time $t$ be $V^{*}_t = V_{t}(K^{*}, T_2-t)$. As already defined in Section \ref{Superior Predictive Analysis}, let $\Delta V^{*}_{t}$ be the realised PnL of target option, whereas, $\Delta \mathcal{G}_{m,t}$ be the realised model PnL, where $m$ $=$ $static$ for static hedge model (constant volatility linear model) and $m$ $=$ $dyn$ for dynamic hedge model. The usage of implied volatility surface (by the implied volatility construction methodology, as explained in Section \ref{Implied Volatility Surface Construction}) ensures that the realized closing price is equal to the model price discussed in Equation \ref{BS}. 

\item The marginal PnL with respect to a risk factor is considered as the change in the portfolio or option value caused by the changes in the corresponding risk factor while keeping all other variables constant. In other words, to calculate the marginal PnL for a specific risk factor, we revalue the trade/portfolio using Equation \ref{BS} with the corresponding risk factor as of the valuation date and all other variables (market and trade-specific) as of the previous business day. The difference in the trade or portfolio value is considered the marginal PnL. We can denote the marginal PnL at any time $t$ with respect to spot, implied volatility, and time to maturity as  $\Delta V^{*}_{t}(S)$,  $\Delta V^{*}_{t}(\sigma)$ and $\Delta V^{*}_{t}(\tau)$ respectively  for the target option and $\Delta \mathcal{G}_{static,t}(S)$,  $\Delta \mathcal{G}_{static,t}(\sigma)$ and $\Delta \mathcal{G}_{static,t}(\tau)$ respectively for static hedge portfolio.

\item The risk PnL is defined as the change in the value of option or portfolio with respect to certain risk sensitivity. The different risk sensitivities (option Greeks) considered at time $t$ are: delta ($\delta_t$), vega ($\nu_t$), and theta ($\Theta_t$) which measure the first-order sensitivities of an option's price to changes in underlying asset price, implied volatility, and time decay respectively. Gamma ($\gamma_t$) and volga ($\vartheta_t$) be the second-order sensitivity of an option's price to changes in the underlying asset price and implied volatility respectively. Therefore, by the classic risk-based explain using a Taylor expansion, an arbitrary option PnL ($\Delta V_t$) can be broken into:

\begin{align}\label{Risk-PnL}
\Delta V_t &= \delta_{t-1} \cdot \Delta S_t + \frac{1}{2} \cdot  \gamma_{t-1} \cdot {(\Delta S_t)}^{2} \nonumber \\
& + \nu_{t-1} \cdot \Delta \sigma_t + \frac{1}{2} \cdot   \vartheta_{t-1} \cdot {(\Delta \sigma_t)}^{2} \nonumber \\ 
& + \Theta_{t-1} \cdot \frac{1}{365}+ \varepsilon, \nonumber \\
&= \Delta V_t (\delta) + \Delta V_t (\gamma) + \Delta V_t (\nu) + \Delta V_t (\vartheta) + \Delta V_t (\Theta) + \varepsilon,
\end{align} 

 where, $\Delta S_t$, $\Delta \sigma_t$ is the change in underlying and implied volatility from $t-1$ to $t$, and $\varepsilon$ is the un-explained PnL. In equation \ref{Risk-PnL}, each element of the summation on the right-hand side corresponds to the delta PnL, gamma PnL, vega PnL, volga PnL, and theta PnL, in that exact order, for a given option with full PnL $\delta_t$.  \\

Therefore, the target option PnL, $\Delta V^{*}_{t}$ can be attributed to delta $\Delta V^{*}_t (\delta)$, gamma $\Delta V^{*}_t (\gamma)$, vega $\Delta V^{*}_t (\nu)$, volga $\Delta V^{*}_t (\vartheta),$ theta $\Delta V^{*}_t (\Theta),$ and un-explained PnLs as given by Equation \ref{Risk-PnL}. 

Similarly, the static hedge PnL can be divided into constituent options' risk PnLs. Therefore, to calculate the risk PnL (corresponding to any sensitivity) of a static hedge portfolio, the risk PnL of each individual option in the portfolio is computed and scaled by corresponding portfolio weights and then aggregated over all constituent options.  The full PnL of the static hedge, $\Delta \mathcal{G}_{static,t}$, can be attributed to the risk PnLs: delta  $\Delta \mathcal{G}_{static,t} (\delta)$, gamma $\Delta \mathcal{G}_{static,t} (\gamma)$, vega $\Delta \mathcal{G}_{static,t}(\nu)$, volga $\Delta \mathcal{G}_{static,t} (\vartheta)$ and theta $\Delta \mathcal{G}_{static,t} (\Theta)$ using Equation \ref{Risk-PnL}.

\end{enumerate}

For the dynamic hedge portfolio of target option with delta $\delta^{*}_{t}$, the one day full PnL $\Delta \mathcal{G}_{dyn,t}$ can be attributed as:

\begin{align}\label{dynamic-PnL}
\Delta \mathcal{G}_{dyn,t} &= \delta^{*}_{t-1} \cdot \Delta S + \big(V^{*}_{t-1} - \delta^{*}_{t-1} \cdot S_{t-1} \big) \cdot (e^{\frac{r}{365}} - 1).
\end{align} 

It's important to note that the full PnL of the dynamic delta hedge is solely influenced by the option's delta and is not sensitive to Gamma or Vega.

%\begin{align}\label{Target-Risk-PnL}
%\delta_{target, t} &= \delta^{risk}_{target}(t) + \gamma^{risk}_{target}(t) + \nu^{risk}_{target}(t) + \vartheta^{risk}_{target}(t) + \Theta^{risk}_{target}(t) + \varepsilon_{target}
%\end{align} 

%
% \begin{align}\label{static-PnL}
%\delta_{static, t} &= \delta^{risk}_{static}(t) + \gamma^{risk}_{static}(t) + \nu^{risk}_{static}(t) + \vartheta^{risk}_{static}(t) + \Theta^{risk}_{static}(t) + \varepsilon_{static}
%\end{align} 

 We analyzed the PnLs of NIFTY options, specifically the call options with ATM, ITM, and OTM strikes (moneyness of target option as of the begining of month), over a one-year period that includes the Covid period. The results of our analysis are presented in Figure \ref{ATM_NIFTY_Analysis} for the ATM case. The ITM and OTM results can be observed in Appendix \ref{App:PnL Attribution Analysis} Figures \ref{ITM_NIFTY_Analysis} and \ref{OTM_NIFTY_Analysis} respectively. Each subplot in the figures can be interpreted as follows:

\begin{itemize}
\item In subplot (a), static hedge portfolio PnL error is $\Delta V^{*}_{t} - \Delta \mathcal{G}_{static,t}$ and Dynamic hedge PnL error is $\Delta V^{*}_{t} - \Delta \mathcal{G}_{dyn,t}$ $\forall t$. The PnL errors of linear constant volatility static hedge portfolio and dynamic hedge portfolio with respect to target option PnL for the historical period considered is observed. In periods characterized by substantial fluctuations in the NIFTY index, particularly observed during the Covid period, dynamic hedging has exhibited noteworthy spikes in errors. In contrast, static hedging has consistently demonstrated superior performance during these periods.

\item In subplot (b), we analyze the delta PnL $\Delta V^{*}_t (\delta)$ as well as the delta-gamma PnL $\Delta V^{*}_t (\delta) \ + \ \Delta V^{*}_t (\gamma)$, which is the sum of delta PnL and gamma PnL of the target option. We calculate the PnL error by comparing the marginal PnL $\Delta V^{*}_t (S)$ of target option resulting from index movements (with all other factors constant) with the delta PnL and the delta-gamma PnL. In the plot, the delta PnL error is $|\Delta V^{*}_t (S) \ - \ \Delta V^{*}_t (\delta)|$ and delta-gamma PnL error is $|\Delta V^{*}_t (S) \ - \ \Delta V^{*}_t (\delta) \ - \ \Delta V^{*}_t (\gamma)|$, where |.| represents absolute value. Our observations show a comparable error plot to that of subplot (a). As anticipated, the delta PnL error exhibits spikes at points where the dynamic hedge performs inadequately, as the dynamic hedge only covers delta risk. In contrast, the delta-gamma PnL error demonstrates a comparable error plot to that of the static hedge, implying the possibility of static hedging to manage both delta and gamma risks.

\item In subplot (c), to establish the evidence for effectiveness of static hedging to mitigate both delta and gamma risks, we conduct a regression analysis between the delta-gamma PnL of the static hedge portfolio $\Delta \mathcal{G}_{static,t} (\delta) \ + \ \Delta \mathcal{G}_{static,t} (\gamma)$ and the target option $\Delta V^{*}_t (\delta) \ + \ \Delta V^{*}_t (\gamma)$, across all $t$. Our results indicate a remarkably high R-square value of over $99\%$ in all cases (including ATM, ITM, OTM), with a beta coefficient near unity. These findings clearly demonstrate the ability of static hedging to effectively cover both delta and gamma risks.

\item In subplot (d), a regression analysis is conducted to examine the effectiveness of static hedge in capturing vega risk. Specifically, the static hedge vega PnL $\Delta \mathcal{G}_{static,t}(\nu)$ is regressed against the target option vega PnL $\Delta V^{*}_t (\nu)$, revealing that the static hedge strategy only offers partial coverage of the vega risk. One contributing factor to the observed partial coverage of vega risk by the static hedge strategy is believed to be the maturity mismatch between the static hedge and the target option. The intuition behind the observed phenomenon can be attributed to the difference in the vega of the target option and that of the shorter-term portfolio. The former corresponds to a longer maturity, while the latter is determined by a smaller time horizon. The impact of changes in implied volatility on static hedge PnL may require higher-order terms and interaction terms, especially as the dimensionality of the surface increases with the number of constituent options. However, a detailed investigation of these effects is beyond the scope of this study and remains an avenue for future research.
\end{itemize}

From Figures \ref{P2_NIFTY_Linear_ATM_CE_static_vs_dynamic_pnL_error.png}, we identified a few key dates for the NIFTY ATM call target option and focussed the PnL attirbution analysis to the following two cases: 

\begin{itemize}
 \item Case1: Dynamic delta hedging had the worst performance 
 \item Case2:  Static hedging had the worst performance. 
\end{itemize}

 Additionally, we expanded our analysis to include ITM and OTM options for the same dates, noting that the moneyness of the target option was determined based on the beginning of the month rather than on the analysis date. The PnL attribution analysis results are elucidated in Tables \ref{tab:realstatdynPnL}-\ref{PnL Analysis - Change in Time}, delineating full, delta and gamma, vega, and theta PnL explanations respectively.\\

\begin{itemize}

\item {\bf Case1: } The objective is to identify the factors that contributed to the superior performance of the Constant Volatility Linear Lasso Static hedge model over the dynamic hedge model. As shown in the PnL error plot in Figure \ref{P2_NIFTY_Linear_ATM_CE_static_vs_dynamic_pnL_error.png}, we observed two large spikes in the dynamic hedge error during the Covid period on April 7, 2020 and May 4, 2020, while the constant volatility static hedge error remained consistently lower. Table \ref{tab:realstatdynPnL} provides the realized target option PnL ($\Delta V^{*}_{t}$), delta hedge PnL ($ \Delta \mathcal{G}_{dyn,t}$), and static hedge PnL ($ \Delta \mathcal{G}_{static,t}$) corresponding to these dates. It is worth noting that the NIFTY index experienced an $8.76\%$ rise and $5.74\%$ fall on April 7 and May 4, respectively. \\

Firstly, we examine the PnL for the target option, static hedge portfolio, and dynamic hedge portfolio resulting from changes solely in index levels, with other risk factors such as volatility held constant. On both dates, it was observed that relying solely on delta PnL couldn't adequately explain the marginal impact due to index changes. However, when combined with gamma PnL, it explained the majority of the impact. Notably, the gamma PnL as a percentage of marginal PnL due to only index change was quite substantial, accounting for approximately 26\% and 44\% for the NIFTY ATM target option on the two dates considered for the analysis.

As already indicated in the regression analysis, the delta PnL and gamma PnL of the target option were closely aligned with the delta and gamma PnL of the static hedge portfolio in most cases. This suggests the potential effectiveness of static hedge in mitigating gamma risk. In contrast, the dynamic hedge, which captures only delta risk, falls short in mitigating gamma risk. This limitation is identified as a key reason for the poor performance of the hedge on the selected dates for the first case. The marginal PnL due to index change, along with delta and gamma PnL, is provided in Table \ref{tab:deltagamma} for the analysis dates.

Secondly, we look at the remaining change in the PnL explained by other risk factors: implied volatility (which is again not captured by a dynamic hedge) and passage of time (theta risk). We could explain most of the marginal PnL due to change in implied volatility by Vega and Volga PnL, whereas, the marginal PnL due to change in time by Theta PnL for target option and static hedge portfolio individually. If we accommodate the change in the implied volatility and theta factor, the remaining portion of the PnL (apart from delta and gamma exposure) is relatively close between the static hedge portfolio and the target option, thereby causing the full PnL to be closer to the target option PnL when compared with a dynamic hedge.   For a dynamic hedge, the marginal PnL due to the passage of time (contributed by deposit account) is very low as expected, thereby not accommodating the theta risk of the target option. The marginal PnLs due to volatility change (along with vega PnL and Volga PnL) and marginal PnL due to passage of time (and theta PnL) are available in Tables \ref{tab:pnl-iv} and \ref{PnL Analysis - Change in Time} respectively. 

\item {\bf Case2:} We focused on the worst-performing case of the static hedge model (March 30, 2020) for the NIFTY ATM call option. Notably, the worst-performance static hedge error was not as high as the worst-performance dynamic hedge error. For the target option, we found that the delta PnL was nearly equivalent to the full PnL, while the vega PnL and gamma PnL were low and largely offset by the theta PnL. However, for the static hedge model, the delta PnL and gamma PnL were closer to the target option, but the relatively lower vega PnL and larger negative theta PnL compared to target option resulted in the difference in full PnL.

\end{itemize}

\begin{figure}[!htb]%
 \centering
 \subfloat[]{\includegraphics[width=0.5\textwidth, height=2in]{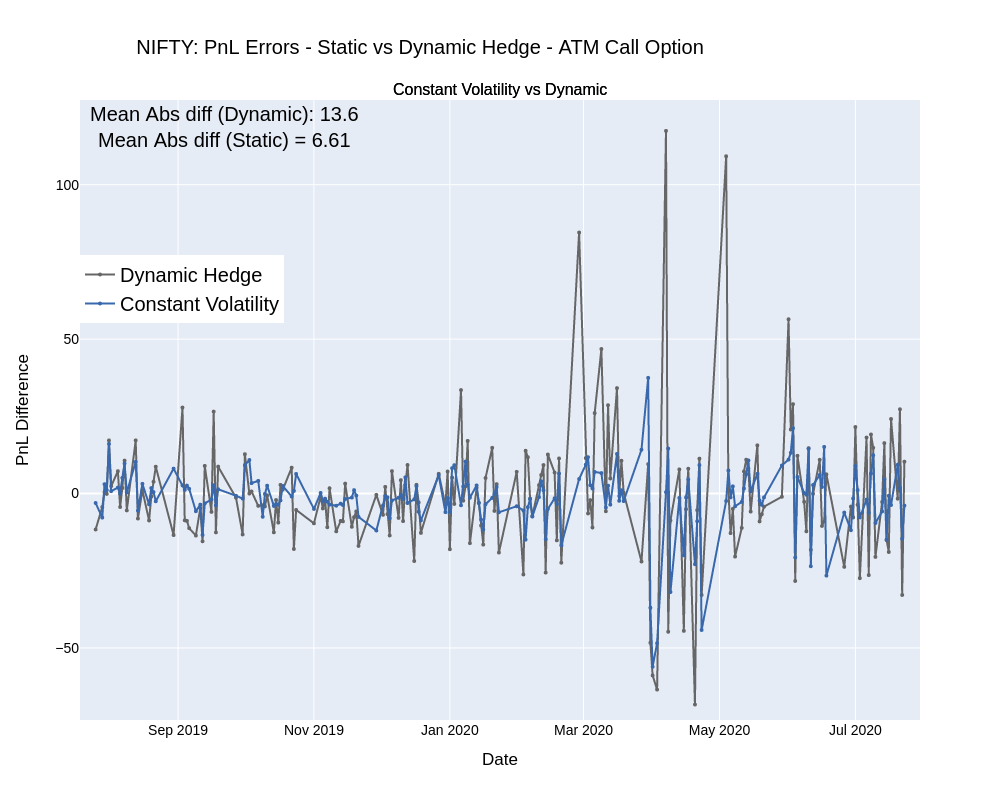}\label{P2_analysis_section_NIFTY_Linear_ATM_CE_static_vs_dynamic_pnL_error.png}}%
 \subfloat[]{\includegraphics[width=0.5\textwidth, height=2in]{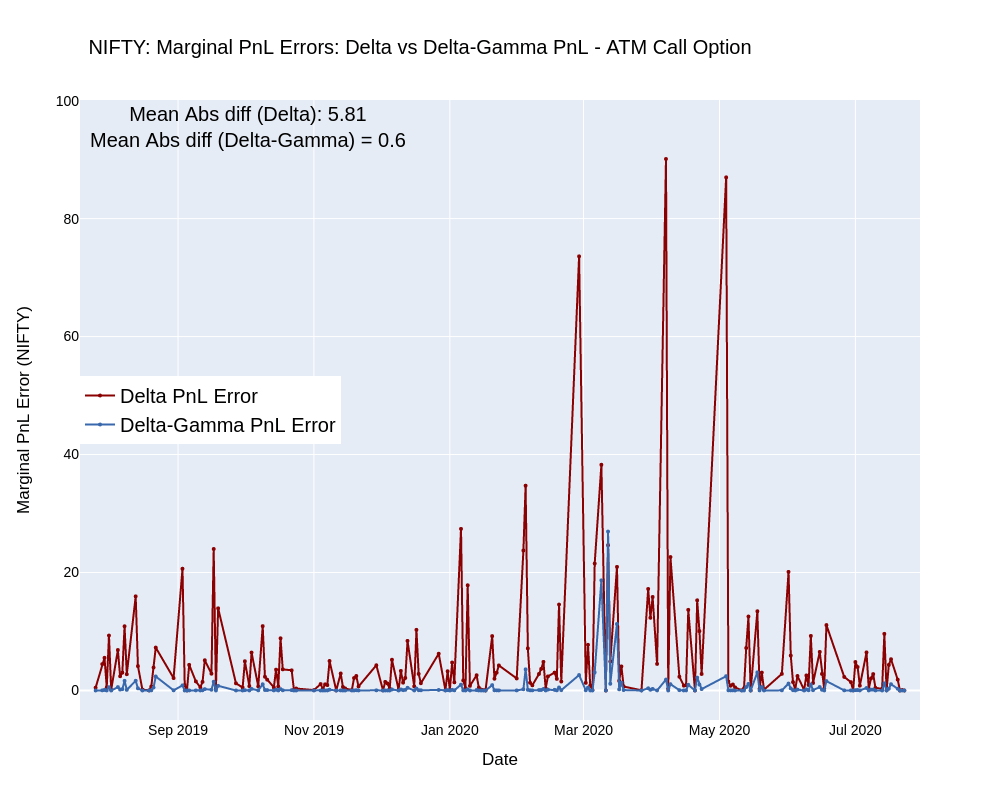}\label{P5_analysis_section_NIFTY_ATM_CE_delta_vs_deltagamma.png}}\\
 \subfloat[]{\includegraphics[width=0.5\textwidth, height=2in]{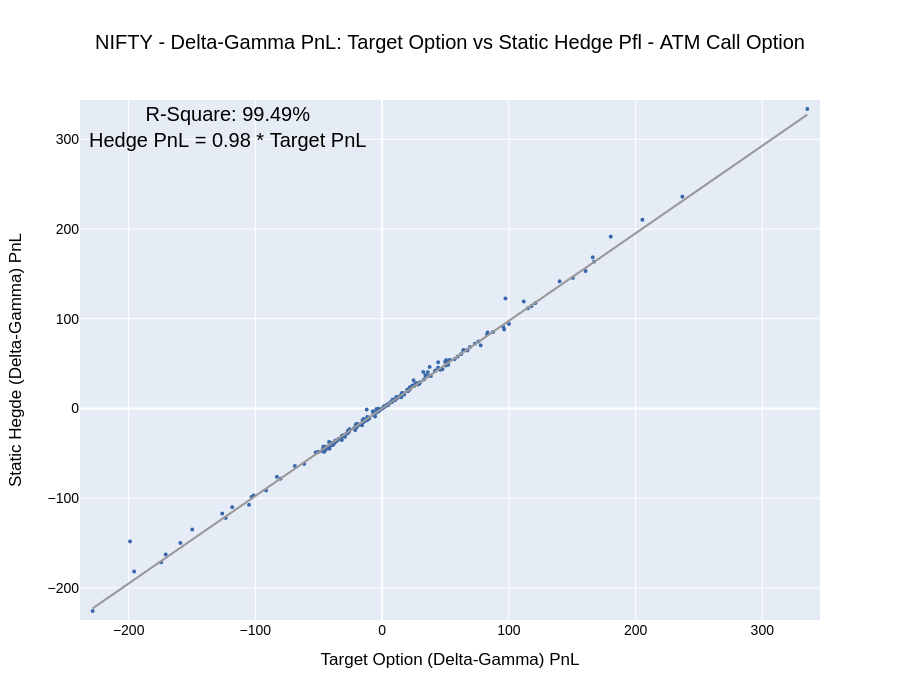}\label{P5_analysis_section_NIFTY_ATM_CE_deltagamma_regression.png}}%
 \subfloat[]{\includegraphics[width=0.5\textwidth, height=2in]{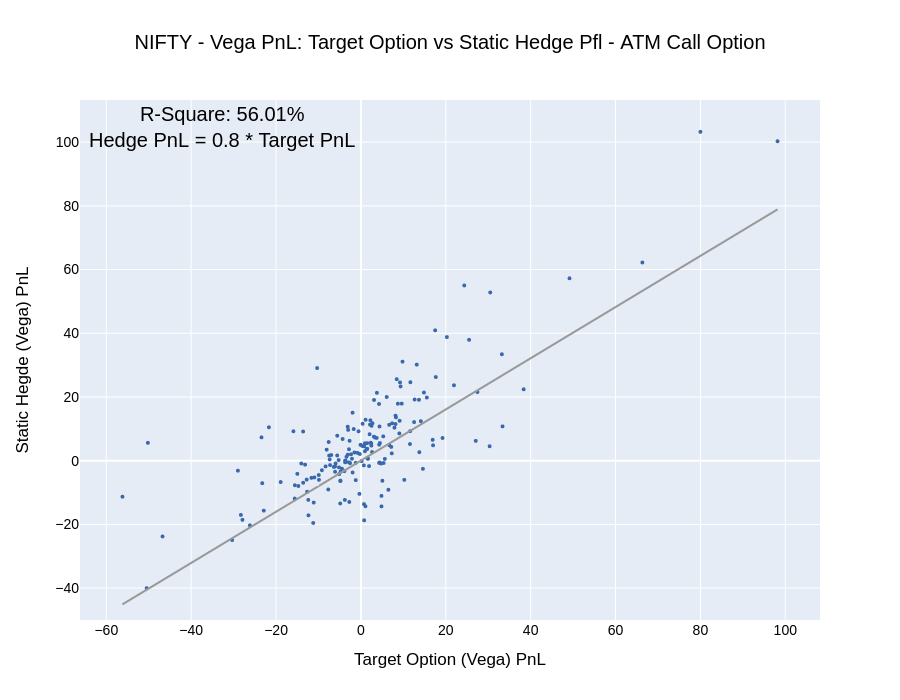}\label{P5_analysis_section_NIFTY_ATM_CE_vega_regression.png}}%
 \caption{\footnotesize The figure illustrates the PnL Attribution Analysis to identify the factors that explain superior performance of Lasso static hedging in NIFTY ATM CALL options. In subplots (a) and (b), the x-axis correspond to historical dates and y-axis correspond to PnL error or difference with respect to target option PnL. In subplots (c) and (d), the x-axis correspond to risk PnL of target option and y-axis correspond to risk PnL of static hedge portfolio. In subplot (a), the historical PnL error of constant volatility linear static hedge model and dynamic hedge model. The PnL error of a hedge model is calculated as the difference between realised target option PnL and realised hedge model portfolio PnL. In subplot (b), we compare the absolute differences of the target option PnL with respect to two risk PnLs of the target option: delta PnL, combined delta and gamma PnL. In subplot (c), we regress the delta-gamma PnL of target option and static hedge portfolio. In subplot (d), we regress the vega PnL of target option and static hedge portfolio.}%
 \label{ATM_NIFTY_Analysis}%
\end{figure}

% Table generated by Excel2LaTeX from sheet 'doc'
\begin{table}[!htb]
  \centering
\resizebox{0.85\textwidth}{!}{%
    \begin{tabular}{|c|p{3em}|c|c|c|c|c|}
    \toprule
    \multicolumn{1}{|p{5.135em}|}{\textbf{Date}} & \textbf{Money -ness} & \multicolumn{1}{p{5.455em}|}{\textbf{Target Option  PnL}} & \multicolumn{1}{p{5.91em}|}{\textbf{Static Hedg PnL}} & \multicolumn{1}{p{6.775em}|}{\textbf{Delta Hedge PnL}} & \multicolumn{1}{p{7.275em}|}{\textbf{Static Hedge Error}} & \multicolumn{1}{p{6.865em}|}{\textbf{Delta Hedge Error}} \\
    \midrule
    \multirow{3}[6]{*}{\textbf{7 April 2020}} & ATM      & 360.00   & 363.55   & 246.15   & -3.55    & 113.85 \\
\cmidrule{2-7}             & ITM      & 519.40   & 513.80   & 398.71   & 5.60     & 120.69 \\
\cmidrule{2-7}             & OTM      & 106.15   & 144.57   & 71.83    & -38.42   & 34.32 \\
    \midrule
    \multirow{3}[6]{*}{\textbf{4 May 2020}} & ATM      & -163.85  & -158.96  & -290.39  & -4.89    & 126.54 \\
\cmidrule{2-7}             & ITM      & -362.70  & -362.15  & -516.73  & -0.55    & 154.03 \\
\cmidrule{2-7}             & OTM      & -7.60    & -12.64   & -34.40   & 5.04     & 26.80 \\
    \midrule
    \multirow{3}[6]{*}{\textbf{30 March 2020}} & ATM      & -200.00  & -231.44  & -214.33  & 31.44    & 14.33 \\
\cmidrule{2-7}             & ITM      & -277.85  & -267.21  & -262.55  & -10.64   & -15.30 \\
\cmidrule{2-7}             & OTM      & -79.60   & -145.00  & -123.68  & 65.40    & 44.08 \\
    \bottomrule
    \end{tabular}%
}
  \caption{\footnotesize This table presents the PnL of target option, static hedge portfolio and dynamic hedge portfolio on 7 APR, 4 MAY and 30 MAR 2020 for ATM, ITM and OTM NIFTY call options. Additionally, the table includes the PnL error, calculated as the difference between the target option PnL and the hedge option PnL for both static and dynamic hedging models.}
  \label{tab:realstatdynPnL}
\end{table}

\begin{table}[!htb]
  \centering
\resizebox{0.85\textwidth}{!}{%
    \begin{tabular}{|c|p{3em}|c|c|c|c|c|c|}
    \toprule
    \multicolumn{1}{|c|}{\multirow{3}[6]{*}{\textbf{Date}}} & \textbf{Money -ness} & \multicolumn{6}{p{30.13em}|}{\textbf{Change in Index Level}} \\
\cmidrule{3-8}             & \multicolumn{1}{c|}{} & \multicolumn{3}{p{14.68em}|}{\textbf{Target Option}} & \multicolumn{3}{p{15.45em}|}{\textbf{Static Hedge Portfolio}} \\
\cmidrule{3-8}             & \multicolumn{1}{c|}{} & \multicolumn{1}{p{5.455em}|}{\textbf{Marginal PnL}} & \multicolumn{1}{p{4.68em}|}{\textbf{Delta PnL}} & \multicolumn{1}{p{4.545em}|}{\textbf{Gamma PnL}} & \multicolumn{1}{p{5.635em}|}{\textbf{Marginal PnL}} & \multicolumn{1}{p{5.135em}|}{\textbf{Delta PnL}} & \multicolumn{1}{p{4.68em}|}{\textbf{Gamma PnL}} \\
    \midrule
    \multirow{3}[6]{*}{\textbf{7 April 2020}} & ATM      & 337.39   & 247.28   & 88.25    & 337.47   & 240.12   & 93.66 \\
\cmidrule{2-8}             & ITM      & 476.74   & 400.49   & 83.11    & 466.84   & 373.05   & 99.27 \\
\cmidrule{2-8}             & OTM      & 131.36   & 72.17    & 46.36    & 144.33   & 79.01    & 51.18 \\
    \midrule
    \multirow{3}[6]{*}{\textbf{4 May 2020}} & ATM      & -201.32  & -288.31  & 89.42    & -168.81  & -272.10  & 123.81 \\
\cmidrule{2-8}             & ITM      & -461.48  & -513.22  & 37.00    & -376.93  & -477.19  & 114.04 \\
\cmidrule{2-8}             & OTM      & -15.25   & -34.15   & 29.01    & -14.30   & -35.07   & 32.83 \\
    \midrule
    \multirow{3}[6]{*}{\textbf{30 March 2020}} & ATM      & -195.30  & -212.52  & 16.84    & -181.98  & -200.48  & 18.64 \\
\cmidrule{2-8}             & ITM      & -246.48  & -260.43  & 13.35    & -229.25  & -245.97  & 16.57 \\
\cmidrule{2-8}             & OTM      & -105.00  & -122.57  & 18.14    & -103.06  & -120.21  & 18.01 \\
    \bottomrule
    \end{tabular}%
}
  \caption{\footnotesize The table presents the marginal PnL analysis with respect to Index Levels. The table presents the marginal PnL due to index levels along with risk PnLs with respect to delta and gamma for both target option and static hedge portfolio.}
  \label{tab:deltagamma}
\end{table}

% Table generated by Excel2LaTeX from sheet 'doc'
\begin{table}[!htb]
  \centering
\resizebox{0.85\textwidth}{!}{%
    \begin{tabular}{|c|p{3em}|c|c|c|c|c|c|}
    \toprule
    \multicolumn{1}{|c|}{\multirow{3}[6]{*}{\textbf{Date}}} & \textbf{Money -ness} & \multicolumn{6}{p{30.13em}|}{\textbf{Change in Implied Volatility}} \\
\cmidrule{3-8}             & \multicolumn{1}{c|}{} & \multicolumn{3}{p{14.68em}|}{\textbf{Target Option}} & \multicolumn{3}{p{15.45em}|}{\textbf{Static Hedge Portfolio}} \\
\cmidrule{3-8}             & \multicolumn{1}{c|}{} & \multicolumn{1}{p{5.455em}|}{\textbf{Marginal PnL}} & \multicolumn{1}{p{4.68em}|}{\textbf{Vega PnL}} & \multicolumn{1}{p{4.545em}|}{\textbf{Volga PnL}} & \multicolumn{1}{p{5.635em}|}{\textbf{Marginal PnL}} & \multicolumn{1}{p{5.135em}|}{\textbf{Vega PnL}} & \multicolumn{1}{p{4.68em}|}{\textbf{Volga PnL}} \\
    \midrule
    \multirow{3}[6]{*}{\textbf{7 April 2020}} & ATM      & 56.74    & 56.03    & 1.62     & 153.16   & 135.74   & 26.95 \\
\cmidrule{2-8}             & ITM      & 95.89    & 95.88    & 0.06     & 200.13   & 174.44   & 37.56 \\
\cmidrule{2-8}             & OTM      & 1.49     & 1.48     & 0.02     & 64.13    & 55.66    & 12.42 \\
    \midrule
    \multirow{3}[6]{*}{\textbf{4 May 2020}} & ATM      & 80.00    & 80.02    & -0.01    & 93.95    & 76.21    & 27.02 \\
\cmidrule{2-8}             & ITM      & 84.84    & 63.54    & 27.16    & 60.96    & 51.17    & 15.78 \\
\cmidrule{2-8}             & OTM      & 44.82    & 31.00    & 15.28    & 40.49    & 25.43    & 20.08 \\
    \midrule
    \multirow{3}[6]{*}{\textbf{30 March 2020}} & ATM      & 27.47    & 27.47    & 0.00     & 22.54    & 14.97    & 6.58 \\
\cmidrule{2-8}             & ITM      & 1.58     & 1.58     & 0.00     & 16.59    & 11.46    & 4.53 \\
\cmidrule{2-8}             & OTM      & 57.85    & 57.02    & 0.91     & 35.20    & 21.74    & 11.34 \\
    \bottomrule
    \end{tabular}%
}
  \caption{\footnotesize The table presents the marginal PnL analysis with respect to implied volatility. The table presents the marginal PnL due to implied volatility along with risk PnLs with respect to vega and volga for both target option and static hedge portfolio. }
  \label{tab:pnl-iv}%
\end{table}%

% Table generated by Excel2LaTeX from sheet 'doc'
\begin{table}[!htb]
  \centering
\resizebox{0.85\textwidth}{!}{%
    \begin{tabular}{|c|p{3em}|c|c|c|c|c|}
    \toprule
    \multicolumn{1}{|c|}{\multirow{3}[6]{*}{\textbf{Date}}} & \textbf{Money -ness} & \multicolumn{5}{p{25.41em}|}{\textbf{Passage of Time}} \\
\cmidrule{3-7}             & \multicolumn{1}{c|}{} & \multicolumn{2}{p{10.135em}|}{\textbf{Target Option}} & \multicolumn{2}{p{10em}|}{\textbf{Static Hedge Portfolio}} & \multicolumn{1}{p{5.275em}|}{\textbf{Dynamic Hedge}} \\
\cmidrule{3-7}             & \multicolumn{1}{c|}{} & \multicolumn{1}{p{5.09em}|}{\textbf{Marginal PnL}} & \multicolumn{1}{p{5.045em}|}{\textbf{Theta PnL}} & \multicolumn{1}{p{5.275em}|}{\textbf{Marginal PnL}} & \multicolumn{1}{p{4.725em}|}{\textbf{Theta PnL}} & \multicolumn{1}{l|}{\textbf{Marginal PnL}} \\
    \midrule
    \multirow{3}[6]{*}{\textbf{7 April 2020}} & ATM      & -30.83   & -29.97   & -24.46   & -21.17   & -1.14 \\
\cmidrule{2-7}             & ITM      & -37.82   & -36.42   & -22.53   & -15.43   & -1.77 \\
\cmidrule{2-7}             & OTM      & -12.55   & -13.03   & -13.34   & -13.62   & -0.34 \\
    \midrule
    \multirow{3}[6]{*}{\textbf{4 May 2020}} & ATM      & -23.24   & -22.46   & -14.05   & -13.11   & -2.08 \\
\cmidrule{2-7}             & ITM      & -12.05   & -12.21   & 1.56     & 7.81     & -3.51 \\
\cmidrule{2-7}             & OTM      & -5.51    & -5.90    & -5.45    & -6.23    & -0.25 \\
    \midrule
    \multirow{3}[6]{*}{\textbf{30 March 2020}} & ATM      & -31.33   & -30.64   & -55.88   & -49.94   & -1.82 \\
\cmidrule{2-7}             & ITM      & -32.03   & -31.41   & -44.78   & -36.83   & -2.11 \\
\cmidrule{2-7}             & OTM      & -23.71   & -23.37   & -53.94   & -50.33   & -1.11 \\
    \bottomrule
    \end{tabular}%
}
  \caption{\footnotesize The table presents the marginal PnL analysis with respect to change in time. The table presents the marginal PnL due to change in time along with risk PnL with respect to theta for  target option, dynamic hedge and static hedge portfolio.}
  \label{PnL Analysis - Change in Time}%
\end{table}%

\section{Conclusions}

Based on the empirical analysis and SPA tests, we conclude that the Constant Volatility Linear Lasso regression-based static hedging tends to perform better than dynamic hedging across all trade characteristics and market scenarios considered. We also observe that the superiority of the static hedging model over the dynamic hedging model is relatively more evident during highly volatile market conditions (leading to high vrga exposure) or when there are jumps in underlying (leading to high gamma exposure) like the Covid period. This is mainly because a static hedge has the potential to capture delta risk, gamma risk, and partially theta risk and vega risk, whereas the dynamic hedge could capture only delta risk.  Further, the usage of Lasso regression-based static hedging has clear benefits of better interpretability of the model, generation of smaller hedge portfolio as insignificant coefficients are pushed to zero, and efficient runtime. The average runtime on an Intel Xeon Gold 6130 processor is $14.2$ seconds for constructing the static hedge portfolio for the entire historical period of one year. 

\clearpage

\appendix

\section{Benchmark of Lasso Static Hedge with Dynamic delta hedge}
\label{App:Benchmark of Lasso Static Hedge with Dynamic delta hedge}

% Table generated by Excel2LaTeX from sheet 'Results Tables'
\begin{table}[!htb]
  \centering
\resizebox{0.85\textwidth}{!}{%
    \begin{tabular}{|c|p{3.3em}|p{3em}|c|c|c|c|c|c|}
    \toprule
    \multicolumn{9}{|c|}{\textbf{Dynamic Hedging Comparisons  P-Values}} \\
    \midrule
    \multicolumn{9}{|c|}{\textbf{Benchmark: Dynamic Linear Model}} \\
    \midrule
    \multicolumn{1}{|c|}{\multirow{3}[4]{*}{\textbf{Index}}} & {\multirow{3}[4]{*}{\parbox{1cm}{\textbf{Option \\ Type  }}}} & \multirow{3}[4]{*}{\parbox{1cm}{\textbf{Money \\ ness}}} & \multicolumn{3}{c|}{\textbf{Absolute Error Loss Function}} & \multicolumn{3}{c|}{\textbf{Squared Error Loss Function}} \\
\cmidrule{4-9}             &  &         & \textbf{SPA } & \textbf{SPA } & \textbf{SPA } & \textbf{SPA } & \textbf{SPA } & \textbf{SPA } \\
\cmidrule{4-9}             &  &        & \textbf{Lower} & \textbf{Consistent} & \textbf{ Upper} & \textbf{Lower} & \textbf{Consistent} & \textbf{Upper} \\
    \midrule
    \multicolumn{1}{|c|}{\multirow{6}[12]{*}{NIFTY}} & \multicolumn{1}{c|}{\multirow{3}[6]{*}{Call}} & ATM      & 0.16     & 0.62     & 0.71     & 0.17     & 0.65     & 0.72 \\
\cmidrule{3-9}             &          & ITM      & 0.56     & 0.76     & 0.76     & 0.35     & 0.75     & 0.75 \\
\cmidrule{3-9}             &          & OTM      & 0.54     & 0.66     & 0.66     & 0.31     & 0.68     & 0.68 \\
\cmidrule{2-9}             & \multicolumn{1}{c|}{\multirow{3}[6]{*}{Put}} & ATM      & 0.71     & 0.91     & 0.91     & 0.34     & 0.80     & 0.80 \\
\cmidrule{3-9}             &          & ITM      & 0.19     & 0.62     & 0.62     & 0.23     & 0.59     & 0.59 \\
\cmidrule{3-9}             &          & OTM      & 0.14     & 0.14     & 0.14     & 0.41     & 0.65     & 0.65 \\
    \midrule
    \multicolumn{1}{|c|}{\multirow{6}[12]{*}{BANKNIFTY}} & \multicolumn{1}{c|}{\multirow{3}[6]{*}{Call}} & ATM      & 0.51     & 0.71     & 0.71     & 0.39     & 0.68     & 0.68 \\
\cmidrule{3-9}             &          & ITM      & 0.59     & 0.92     & 0.98     & 0.56     & 0.98     & 0.98 \\
\cmidrule{3-9}             &          & OTM      & 0.45     & 0.45     & 0.45     & 0.65     & 0.97     & 0.97 \\
\cmidrule{2-9}             & \multicolumn{1}{c|}{\multirow{3}[6]{*}{Put}} & ATM      & 0.56     & 0.81     & 0.81     & 0.26     & 0.70     & 0.70 \\
\cmidrule{3-9}             &          & ITM      & 0.42     & 0.56     & 0.83     & 0.35     & 0.52     & 0.83 \\
\cmidrule{3-9}             &          & OTM      & 0.27     & 0.45     & 0.45     & 0.58     & 0.89     & 0.89 \\
    \bottomrule
    \end{tabular}%
}
  \caption{\footnotesize SPA Comparison among Dynamic Hedge Model alternatives based on implied volatility surface interpolation schemes. The benchmark model is dynamic hedge model devicing linear interpolation in implied volatility surface. The test is conducted with NIFTY and BANKNIFTY call and put options across ATM, ITM and OTM moneyness levels employing two loss functions: Absolute and Squared Erros loss function. In the table, each value corresponds to the p-value associated with a specific combination of option characteristics and loss function. The test is performed at 95\% confidence level.}
  \label{tab:dynamic_winner}%
\end{table}%

% Table generated by Excel2LaTeX from sheet 'Results Tables'
\begin{table}[!htb]
  \centering
\resizebox{0.85\textwidth}{!}{%
    \begin{tabular}{|c|p{3.3em}|p{3em}|c|c|c|c|c|c|}
    \toprule
    \multicolumn{9}{|c|}{\textbf{SPA (P-Values) : Lasso Static Hedge vs Dynamic Hedge}} \\
    \midrule
    \multicolumn{9}{|c|}{\textbf{Benchmark: Constant Volatility Linear Static Hedge Model}} \\
    \midrule

    \multicolumn{1}{|c|}{\multirow{3}[4]{*}{\textbf{Index}}} & {\multirow{3}[4]{*}{\parbox{1cm}{\textbf{Option \\ Type  }}}} & \multirow{3}[4]{*}{\parbox{1cm}{\textbf{Money \\ ness}}} & \multicolumn{3}{c|}{\textbf{Absolute Error Loss Function}} & \multicolumn{3}{c|}{\textbf{Squared Error Loss Function}} \\
\cmidrule{4-9}             &  &         & \textbf{SPA } & \textbf{SPA } & \textbf{SPA } & \textbf{SPA } & \textbf{SPA } & \textbf{SPA } \\
\cmidrule{4-9}             &  &        & \textbf{Lower} & \textbf{Consistent} & \textbf{ Upper} & \textbf{Lower} & \textbf{Consistent} & \textbf{Upper} \\
    \midrule
    \multicolumn{1}{|c|}{\multirow{6}[12]{*}{NIFTY}} & \multicolumn{1}{c|}{\multirow{3}[6]{*}{Call}} & ATM      & 0.53     & 0.53     & 1.00     & 0.51     & 0.51     & 0.99 \\
\cmidrule{3-9}             &          & ITM      & 0.52     & 0.52     & 1.00     & 0.53     & 0.53     & 0.99 \\
\cmidrule{3-9}             &          & OTM      & 0.52     & 0.52     & 0.99     & 0.53     & 0.87     & 0.87 \\
\cmidrule{2-9}             & \multicolumn{1}{c|}{\multirow{3}[6]{*}{Put}} & ATM      & 0.53     & 0.53     & 1.00     & 0.52     & 0.52     & 0.98 \\
\cmidrule{3-9}             &          & ITM      & 0.49     & 0.49     & 1.00     & 0.54     & 0.54     & 1.00 \\
\cmidrule{3-9}             &          & OTM      & 0.53     & 0.53     & 0.98     & 0.56     & 0.56     & 0.96 \\
    \midrule
    \multicolumn{1}{|c|}{\multirow{6}[12]{*}{BANKNIFTY}} & \multicolumn{1}{c|}{\multirow{3}[6]{*}{Call}} & ATM      & 0.50     & 0.50     & 1.00     & 0.56     & 0.56     & 0.99 \\
\cmidrule{3-9}             &          & ITM      & 0.52     & 0.52     & 1.00     & 0.56     & 0.56     & 0.98 \\
\cmidrule{3-9}             &          & OTM      & 0.51     & 0.51     & 1.00     & 0.55     & 0.55     & 0.97 \\
\cmidrule{2-9}             & \multicolumn{1}{c|}{\multirow{3}[6]{*}{Put}} & ATM      & 0.50     & 0.50     & 1.00     & 0.48     & 0.69     & 0.69 \\
\cmidrule{3-9}             &          & ITM      & 0.51     & 0.92     & 0.92     & 0.32     & 0.32     & 0.32 \\
\cmidrule{3-9}             &          & OTM      & 0.53     & 0.53     & 0.99     & 0.55     & 0.92     & 0.92 \\
    \bottomrule
    \end{tabular}%
}
  \caption{\footnotesize SPA comparison among Constant Linear Static Hedge Model and Dynamic Hedge Model with linear interpolation employed for implied volatility surface construction in all the models. The benchmark model is Constant Volatility Linear static hedge model. The test is conducted with NIFTY and BANKNIFTY call/put options across ATM, ITM and OTM moneyness levels employing two loss functions: Absolute and Squared Erros loss function. In the table, each value corresponds to the p-value associated with a specific combination of option characteristics and loss function. The test is performed at 95\% confidence level.}
  \label{tab:stat_dyn_winner}%
\end{table}%

\clearpage

\section{PnL Attribution Analysis}
\label{App:PnL Attribution Analysis}

\begin{figure}[!htb]%
 \centering
 \subfloat[]{\includegraphics[width=0.5\textwidth, height=2in]{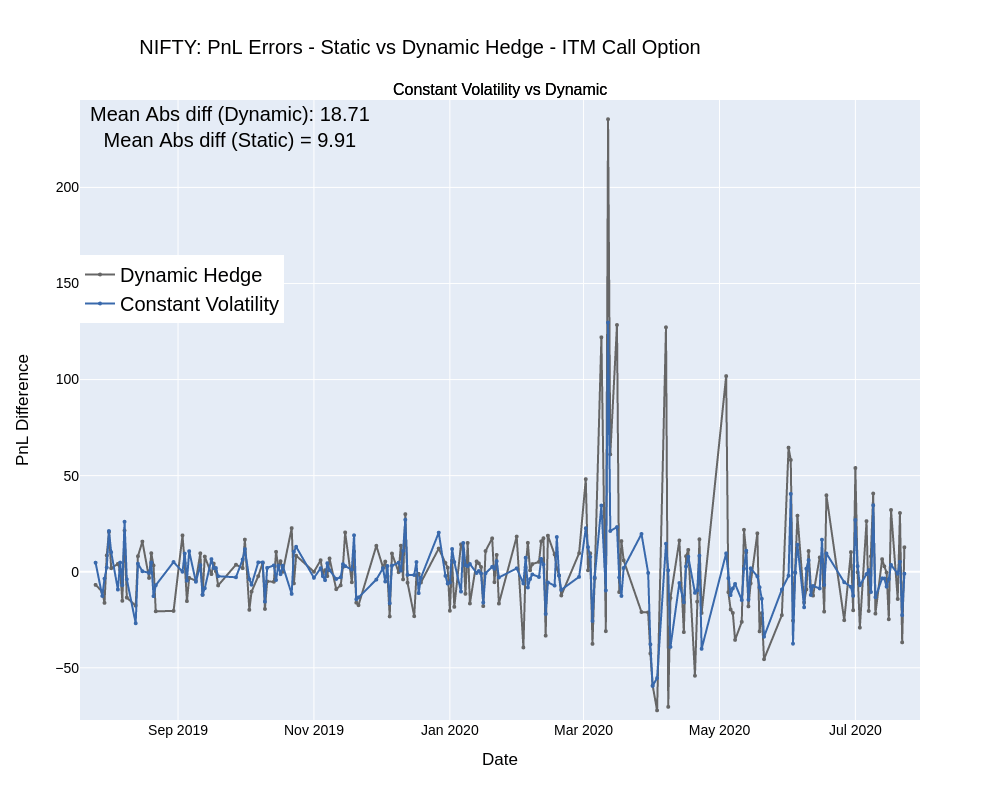}\label{P2_analysis_section_NIFTY_Linear_ITM_CE_static_vs_dynamic_pnL_error.png}}%
 \subfloat[]{\includegraphics[width=0.5\textwidth, height=2in]{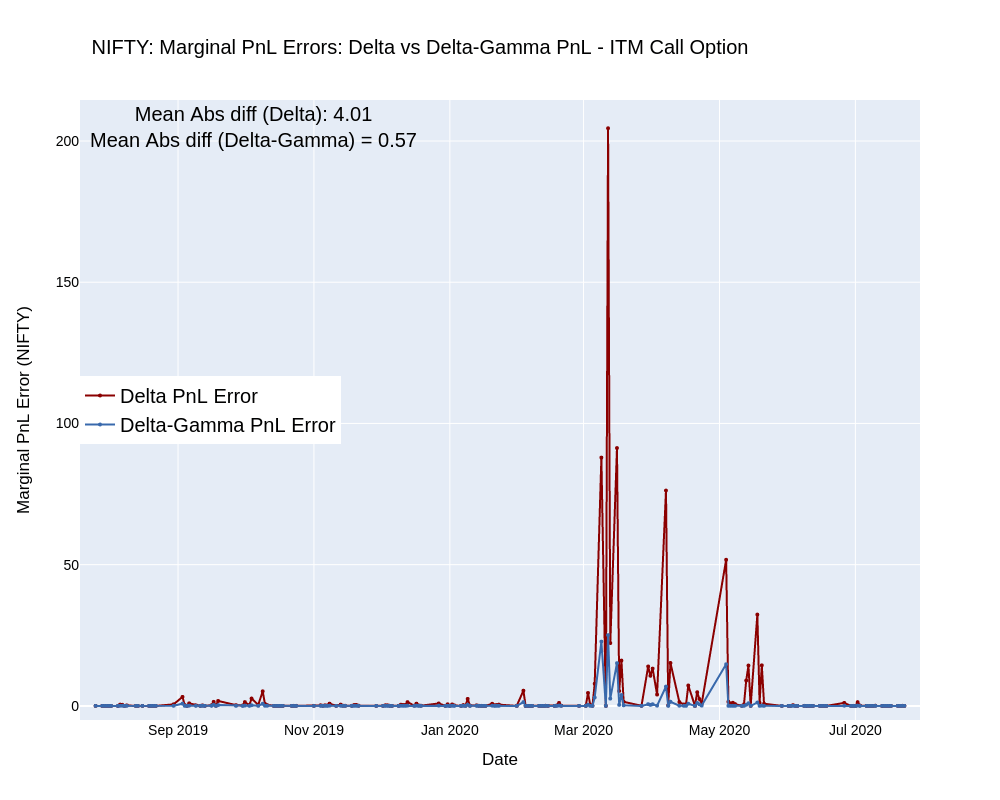}\label{P5_analysis_section_NIFTY_ITM_CE_delta_vs_deltagamma.png}}\\
 \subfloat[]{\includegraphics[width=0.5\textwidth, height=2in]{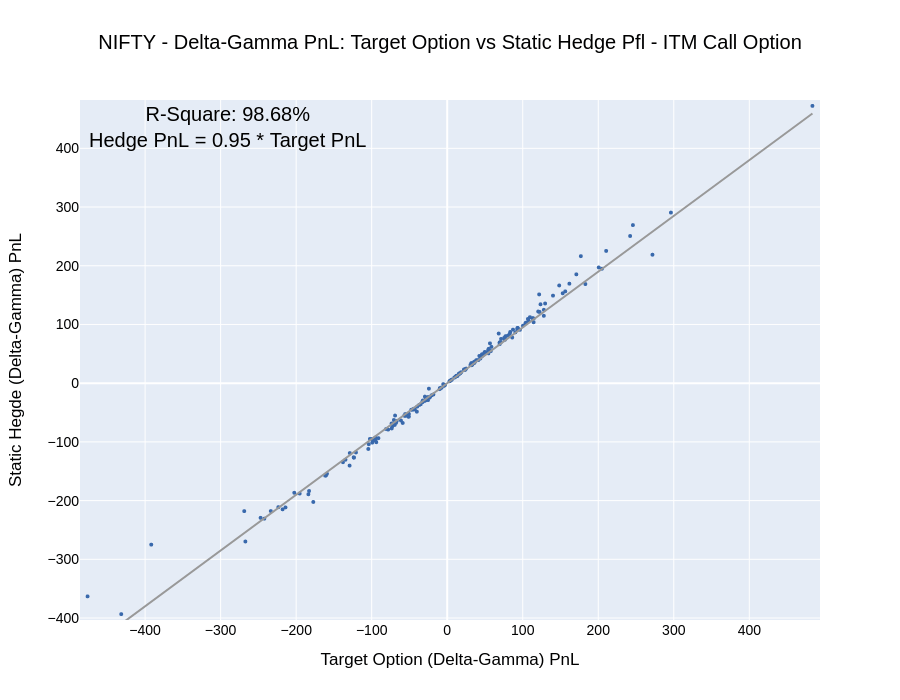}\label{P5_analysis_section_NIFTY_ITM_CE_deltagamma_regression.png}}%
 \subfloat[]{\includegraphics[width=0.5\textwidth, height=2in]{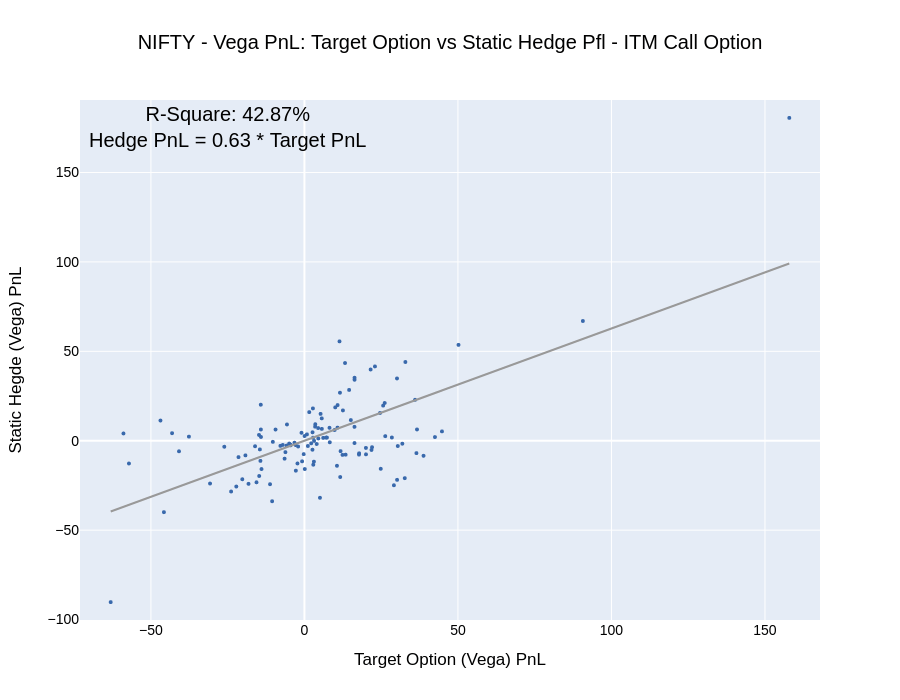}\label{P5_analysis_section_NIFTY_ITM_CE_vega_regression.png}}%
 \caption{\footnotesize The figure illustrates the PnL Attribution Analysis to identify the factors that explain superior performance of Lasso static hedging in NIFTY ITM CALL options. In subplots (a) and (b), the x-axis correspond to historical dates and y-axis correspond to PnL error or difference with respect to target option PnL. In subplots (c) and (d), the x-axis correspond to risk PnL of target option and y-axis correspond to risk PnL of static hedge portfolio. In subplot (a), the historical PnL error of constant volatility linear static hedge model and dynamic hedge model. The PnL error of a hedge model is calculated as the difference between realised target option PnL and realised hedge model portfolio PnL. In subplot (b), we compare the absolute differences of the target option PnL with respect to two risk PnLs of the target option: delta PnL, combined delta and gamma PnL. In subplot (c), we regress the delta-gamma PnL of target option and static hedge portfolio. In subplot (d), we regress the vega PnL of target option and static hedge portfolio.}%
 \label{ITM_NIFTY_Analysis}%
\end{figure}

\begin{figure}[!htb]%
 \centering
 \subfloat[]{\includegraphics[width=0.5\textwidth, height=2in]{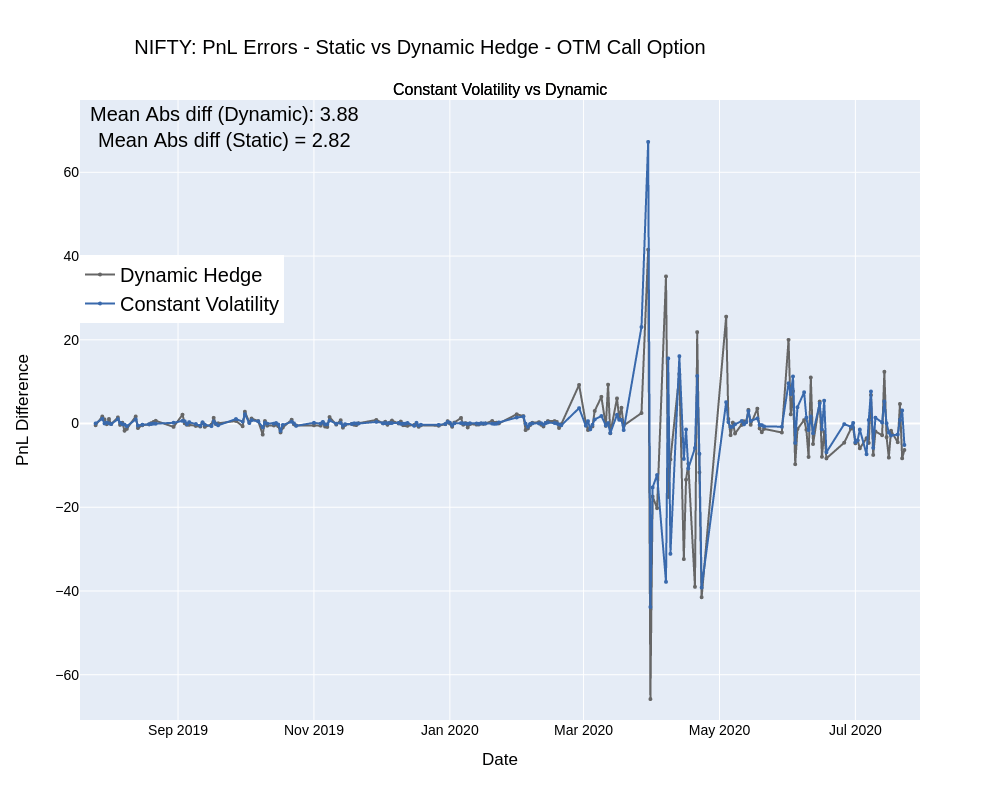}\label{P2_analysis_section_NIFTY_Linear_OTM_CE_static_vs_dynamic_pnL_error.png}}%
 \subfloat[]{\includegraphics[width=0.5\textwidth, height=2in]{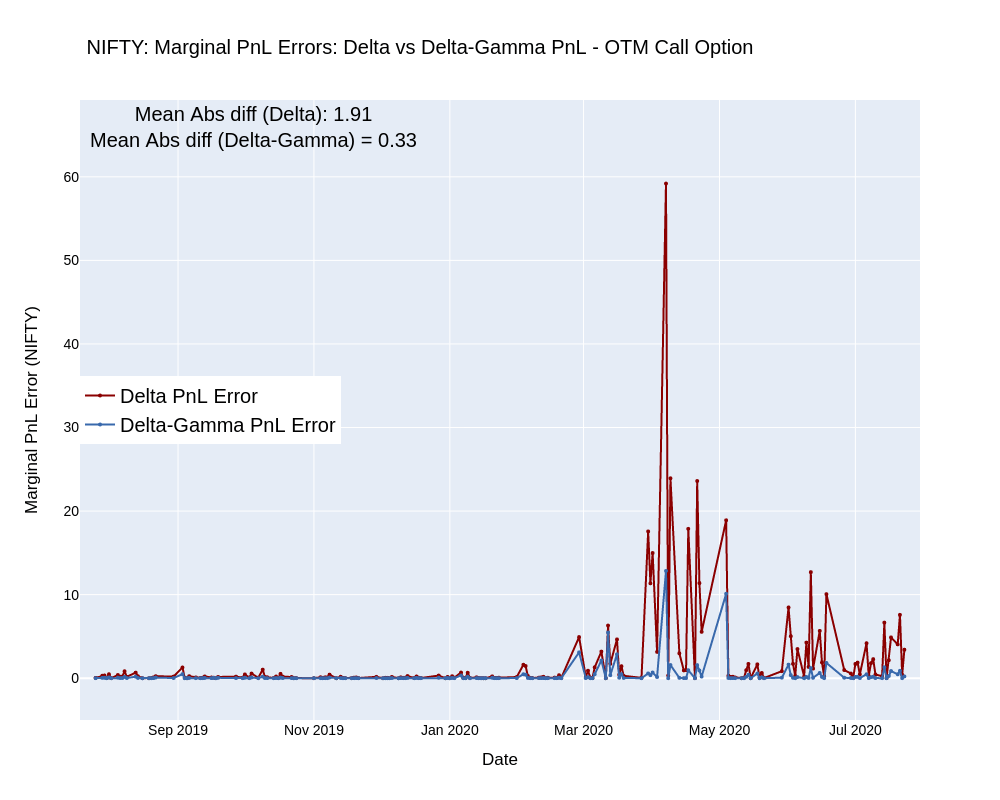}\label{P5_analysis_section_NIFTY_OTM_CE_delta_vs_deltagamma.png}}\\
 \subfloat[]{\includegraphics[width=0.5\textwidth, height=2in]{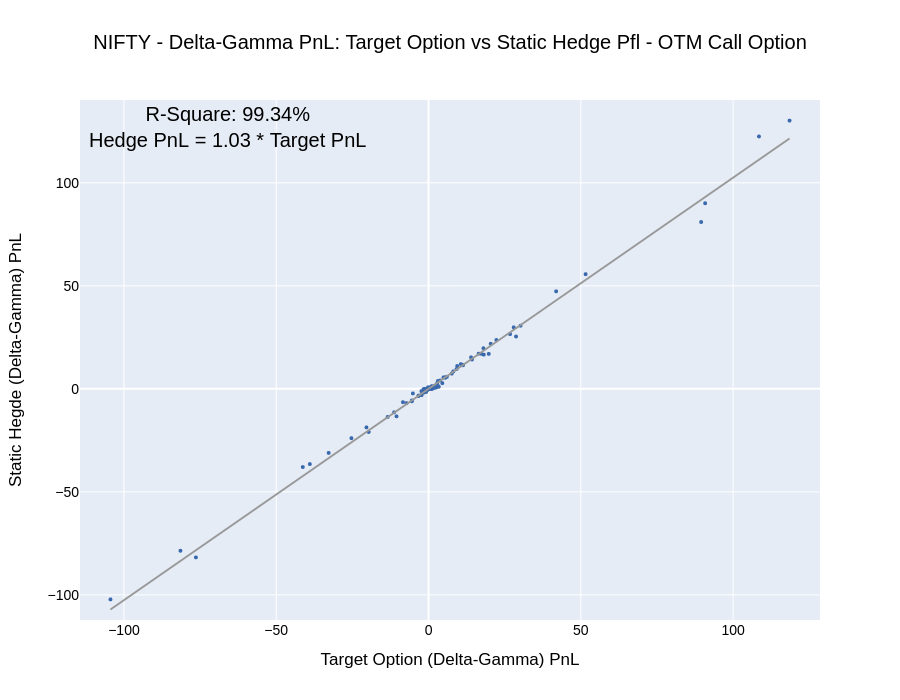}\label{P5_analysis_section_NIFTY_OTM_CE_deltagamma_regression.png}}%
 \subfloat[]{\includegraphics[width=0.5\textwidth, height=2in]{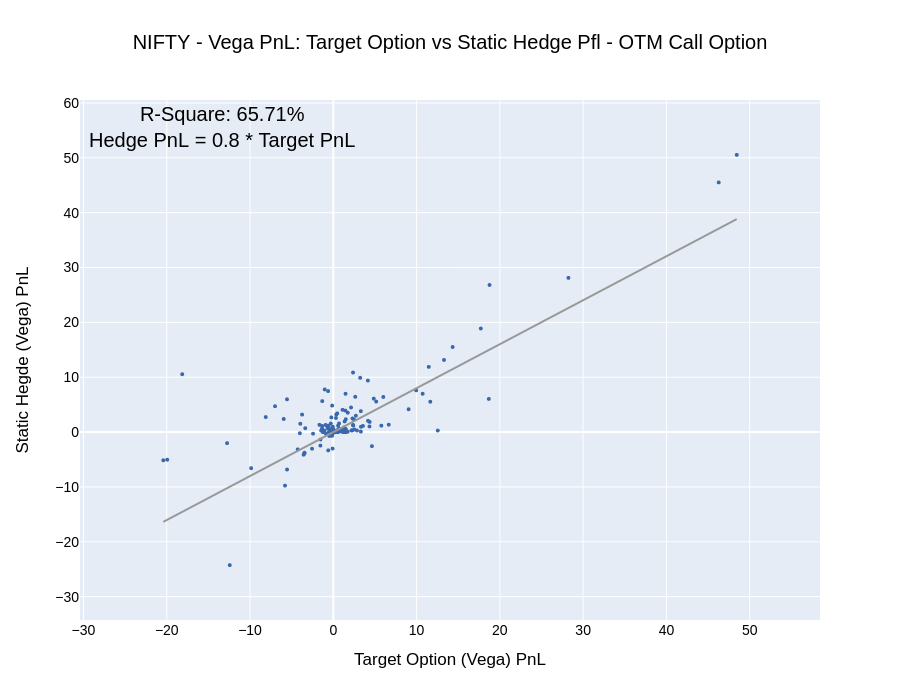}\label{P5_analysis_section_NIFTY_OTM_CE_vega_regression.png}}%
 \caption{\footnotesize The figure illustrates the PnL Attribution Analysis to identify the factors that explain superior performance of Lasso static hedging in NIFTY OTM CALL options. In subplots (a) and (b), the x-axis correspond to historical dates and y-axis correspond to PnL error or difference with respect to target option PnL. In subplots (c) and (d), the x-axis correspond to risk PnL of target option and y-axis correspond to risk PnL of static hedge portfolio. In subplot (a), the historical PnL error of constant volatility linear static hedge model and dynamic hedge model. The PnL error of a hedge model is calculated as the difference between realised target option PnL and realised hedge model portfolio PnL. In subplot (b), we compare the absolute differences of the target option PnL with respect to two risk PnLs of the target option: delta PnL, combined delta and gamma PnL. In subplot (c), we regress the delta-gamma PnL of target option and static hedge portfolio. In subplot (d), we regress the vega PnL of target option and static hedge portfolio.}%
 \label{OTM_NIFTY_Analysis}%
\end{figure}

\clearpage

\bibliographystyle{unsrtnat} % outcomment this and next line in Case 1
\bibliography{Static_Hedging_empirical_study} % if more than one, comma separated

\begin{thebibliography}{33}
\providecommand{\natexlab}[1]{#1}
\providecommand{\url}[1]{\texttt{#1}}
\expandafter\ifx\csname urlstyle\endcsname\relax
  \providecommand{\doi}[1]{doi: #1}\else
  \providecommand{\doi}{doi: \begingroup \urlstyle{rm}\Url}\fi

\bibitem[Carr and Wu(2014)]{carr2014static}
Peter Carr and Liuren Wu.
\newblock Static hedging of standard options.
\newblock \emph{Journal of Financial Econometrics}, 12\penalty0 (1):\penalty0
  3--46, 2014.

\bibitem[Hansen(2005)]{hansen2005test}
Peter~Reinhard Hansen.
\newblock A test for superior predictive ability.
\newblock \emph{Journal of Business \& Economic Statistics}, 23\penalty0
  (4):\penalty0 365--380, 2005.

\bibitem[Bakshi and Kapadia(2003)]{bakshi2003delta}
Gurdip Bakshi and Nikunj Kapadia.
\newblock Delta-hedged gains and the negative market volatility risk premium.
\newblock \emph{The Review of Financial Studies}, 16\penalty0 (2):\penalty0
  527--566, 2003.

\bibitem[Brown et~al.(2001)Brown, Hobson, and Rogers]{brown2001robust}
Haydyn Brown, David Hobson, and Leonard~CG Rogers.
\newblock Robust hedging of barrier options.
\newblock \emph{Mathematical Finance}, 11\penalty0 (3):\penalty0 285--314,
  2001.

\bibitem[Hobson(1998)]{hobson1998robust}
David~G Hobson.
\newblock Robust hedging of the lookback option.
\newblock \emph{Finance and Stochastics}, 2:\penalty0 329--347, 1998.

\bibitem[Fabozzi et~al.(2016)Fabozzi, Paletta, Stanescu, and
  Tunaru]{fabozzi2016improved}
Frank~J Fabozzi, Tommaso Paletta, Silvia Stanescu, and Radu Tunaru.
\newblock An improved method for pricing and hedging long dated american
  options.
\newblock \emph{European Journal of Operational Research}, 254\penalty0
  (2):\penalty0 656--666, 2016.

\bibitem[Wilmott and Schonbucher(2000)]{wilmott2000feedback}
Paul Wilmott and Philipp~J Schonbucher.
\newblock The feedback effect of hedging in illiquid markets.
\newblock \emph{SIAM Journal on Applied Mathematics}, 61\penalty0 (1):\penalty0
  232--272, 2000.

\bibitem[He et~al.(2006)He, Kennedy, Coleman, Forsyth, Li, and
  Vetzal]{he2006calibration}
Changhong He, J~Shannon Kennedy, Thomas~F Coleman, Peter~A Forsyth, Yuying Li,
  and Kenneth~R Vetzal.
\newblock Calibration and hedging under jump diffusion.
\newblock \emph{Review of Derivatives Research}, 9:\penalty0 1--35, 2006.

\bibitem[Lutkebohmert et~al.(2022)Lutkebohmert, Schmidt, and
  Sester]{lutkebohmert2022robust}
Eva Lutkebohmert, Thorsten Schmidt, and Julian Sester.
\newblock Robust deep hedging.
\newblock \emph{Quantitative Finance}, 22\penalty0 (8):\penalty0 1465--1480,
  2022.

\bibitem[Coleman et~al.(2003)Coleman, Kim, Li, and Verma]{coleman2003dynamic}
Thomas~F Coleman, Yohan Kim, Yuying Li, and Arun Verma.
\newblock Dynamic hedging in a volatile market.
\newblock Technical report, Cornell University, 2003.

\bibitem[Coleman et~al.(2001)Coleman, Li, and Verma]{coleman2001reconstructing}
Thomas~F Coleman, Yuying Li, and Arun Verma.
\newblock Reconstructing the unknown local volatility function.
\newblock In \emph{Quantitative Analysis In Financial Markets: Collected Papers
  of the New York University Mathematical Finance Seminar (Volume II)}, pages
  192--215. World Scientific, 2001.

\bibitem[Kennedy et~al.(2009)Kennedy, Forsyth, and Vetzal]{kennedy2009dynamic}
J~Shannon Kennedy, Peter~A Forsyth, and Kenneth~R Vetzal.
\newblock Dynamic hedging under jump diffusion with transaction costs.
\newblock \emph{Operations Research}, 57\penalty0 (3):\penalty0 541--559, 2009.

\bibitem[Breeden and Litzenberger(1978)]{breeden1978prices}
Douglas~T Breeden and Robert~H Litzenberger.
\newblock Prices of state-contingent claims implicit in option prices.
\newblock \emph{Journal of business}, pages 621--651, 1978.

\bibitem[Green and Jarrow(1987)]{green1987spanning}
Richard~C Green and Robert~A Jarrow.
\newblock Spanning and completeness in markets with contingent claims.
\newblock \emph{Journal of Economic Theory}, 41\penalty0 (1):\penalty0
  202--210, 1987.

\bibitem[Nachman(1988)]{nachman1988spanning}
David~C Nachman.
\newblock Spanning and completeness with options.
\newblock \emph{The review of financial studies}, 1\penalty0 (3):\penalty0
  311--328, 1988.

\bibitem[Carr and Madan(2001)]{carr2001optimal}
Peter Carr and Dilip Madan.
\newblock Optimal positioning in derivative securities.
\newblock 2001.

\bibitem[Carr et~al.(1998)Carr, Ellis, and Gupta]{carr1998static}
Peter Carr, Katrina Ellis, and Vishal Gupta.
\newblock Static hedging of exotic options.
\newblock \emph{The Journal of Finance}, 53\penalty0 (3):\penalty0 1165--1190,
  1998.

\bibitem[Takahashi and Yamazaki(2009)]{takahashi2009new}
Akihiko Takahashi and Akira Yamazaki.
\newblock A new scheme for static hedging of european derivatives under
  stochastic volatility models.
\newblock \emph{Journal of Futures Markets: Futures, Options, and Other
  Derivative Products}, 29\penalty0 (5):\penalty0 397--413, 2009.

\bibitem[Takahashi et~al.(2007)Takahashi, Yamazaki,
  et~al.]{takahashi2007efficient}
Akihiko Takahashi, Akira Yamazaki, et~al.
\newblock Efficient static replication of european options for exponential levy
  models (revised in january 2008, published in" journal of futures markets",
  vol. 29-1, 1-15, 2009.).
\newblock Technical report, Center for Advanced Research in Finance, Faculty of
  Economics, The~…, 2007.

\bibitem[Carr and Lee(2009)]{carr2009put}
Peter Carr and Roger Lee.
\newblock Put-call symmetry: Extensions and applications.
\newblock \emph{Mathematical Finance: An International Journal of Mathematics,
  Statistics and Financial Economics}, 19\penalty0 (4):\penalty0 523--560,
  2009.

\bibitem[Carr and Nadtochiy(2011)]{carr2011static}
Peter Carr and Sergey Nadtochiy.
\newblock Static hedging under time-homogeneous diffusions.
\newblock \emph{SIAM Journal on Financial Mathematics}, 2\penalty0
  (1):\penalty0 794--838, 2011.

\bibitem[Wu and Zhu(2016)]{wu2016simple}
Liuren Wu and Jingyi Zhu.
\newblock Simple robust hedging with nearby contracts.
\newblock \emph{Journal of Financial Econometrics}, 15\penalty0 (1):\penalty0
  1--35, 2016.

\bibitem[Leung and Lorig(2016)]{leung2016optimal}
Tim Leung and Matthew Lorig.
\newblock Optimal static quadratic hedging.
\newblock \emph{Quantitative Finance}, 16\penalty0 (9):\penalty0 1341--1355,
  2016.

\bibitem[Bossu et~al.(2021)Bossu, Carr, and Papanicolaou]{bossu2021functional}
S{\'e}bastien Bossu, Peter Carr, and Andrew Papanicolaou.
\newblock A functional analysis approach to the static replication of european
  options.
\newblock \emph{Quantitative Finance}, 21\penalty0 (4):\penalty0 637--655,
  2021.

\bibitem[Bossu et~al.(2022)Bossu, Carr, and Papanicolaou]{bossu2022static}
S{\'e}bastien Bossu, Peter Carr, and Andrew Papanicolaou.
\newblock Static replication of european standard dispersion options.
\newblock \emph{Quantitative Finance}, pages 1--13, 2022.

\bibitem[Buehler et~al.(2019)Buehler, Gonon, Teichmann, and
  Wood]{buehler2019deep}
Hans Buehler, Lukas Gonon, Josef Teichmann, and Ben Wood.
\newblock Deep hedging.
\newblock \emph{Quantitative Finance}, 19\penalty0 (8):\penalty0 1271--1291,
  2019.

\bibitem[Lokeshwar et~al.(2022)Lokeshwar, Bharadwaj, and
  Jain]{lokeshwar2022explainable}
Vikranth Lokeshwar, Vikram Bharadwaj, and Shashi Jain.
\newblock Explainable neural network for pricing and universal static hedging
  of contingent claims.
\newblock \emph{Applied Mathematics and Computation}, 417:\penalty0 126775,
  2022.

\bibitem[Tibshirani(1996)]{tibshirani1996regression}
Robert Tibshirani.
\newblock Regression shrinkage and selection via the lasso.
\newblock \emph{Journal of the Royal Statistical Society: Series B
  (Methodological)}, 58\penalty0 (1):\penalty0 267--288, 1996.

\bibitem[Homescu(2011)]{homescu2011implied}
Cristian Homescu.
\newblock Implied volatility surface: Construction methodologies and
  characteristics.
\newblock \emph{arXiv preprint arXiv:1107.1834}, 2011.

\bibitem[Brent(1971)]{brent1971algorithm}
Richard~P. Brent.
\newblock An algorithm with guaranteed convergence for finding a zero of a
  function.
\newblock \emph{The computer journal}, 14\penalty0 (4):\penalty0 422--425,
  1971.

\bibitem[Politis and Romano(1994)]{politis1994stationary}
Dimitris~N Politis and Joseph~P Romano.
\newblock The stationary bootstrap.
\newblock \emph{Journal of the American Statistical association}, 89\penalty0
  (428):\penalty0 1303--1313, 1994.

\bibitem[Politis and White(2004)]{politis2004automatic}
Dimitris~N Politis and Halbert White.
\newblock Automatic block-length selection for the dependent bootstrap.
\newblock \emph{Econometric reviews}, 23\penalty0 (1):\penalty0 53--70, 2004.

\bibitem[Gon{\c{c}}alves and de~Jong(2003)]{gonccalves2003consistency}
S{\i}lvia Gon{\c{c}}alves and Robert de~Jong.
\newblock Consistency of the stationary bootstrap under weak moment conditions.
\newblock \emph{Economics Letters}, 81\penalty0 (2):\penalty0 273--278, 2003.

\end{thebibliography}

%%%%%%%%%%%%%%%%%
\end{document}